\documentclass[smallextended]{./jar/svjour3}
 
\usepackage{setspace}
\usepackage{amsmath, amssymb, mathtools}

\usepackage[llparenthesis, rrparenthesis]{stmaryrd}

\usepackage{url}

\usepackage{mathpartir}
\newcommand{\inference}[2]{\inferrule{#1}{#2}}

\usepackage[round,authoryear]{natbib}

\usepackage{color}

\usepackage{listings}
\lstdefinelanguage{Coq}{ 
mathescape=true,
texcl=false, 
morekeywords=[1]{Section, Module, End, Require, Import, Export,
  Variable, Variables, Parameter, Parameters, Axiom, Hypothesis,
  Hypotheses, Notation, Local, Tactic, Reserved, Scope, Open, Close,
  Bind, Delimit, Definition, Let, Ltac, Fixpoint, CoFixpoint, Add,
  Morphism, Relation, Implicit, Arguments, Unset, Contextual,
  Strict, Prenex, Implicits, Inductive, CoInductive, Record,
  Structure, Canonical, Coercion, Context, Class, Global, Instance,
  Program, Infix, Theorem, Lemma, Corollary, Proposition, Fact,
  Remark, Example, Proof, Goal, Save, Qed, Defined, Hint, Resolve,
  Rewrite, View, Search, Show, Print, Printing, All, Eval, Check,
  Projections, inside, outside, Def},
morekeywords=[2]{forall, exists, exists2, fun, fix, cofix, struct,
  match, with, end, as, in, return, let, if, is, then, else, for, of,
  nosimpl, when},
morekeywords=[3]{Type, Prop, Set, true, false, option},
morekeywords=[4]{pose, set, move, case, elim, apply, clear, hnf,
  intro, intros, generalize, rename, pattern, after, destruct,
  induction, using, refine, inversion, injection, rewrite, congr,
  unlock, compute, ring, field, fourier, replace, fold, unfold,
  change, cutrewrite, simpl, have, suff, wlog, suffices, without,
  loss, nat_norm, assert, cut, trivial, revert, bool_congr, nat_congr,
  symmetry, transitivity, auto, split, left, right, autorewrite},
morekeywords=[5]{by, done, exact, reflexivity, tauto, romega, omega,
  assumption, solve, contradiction, discriminate},
morekeywords=[6]{do, last, first, try, idtac, repeat},
morecomment=[s]{(*}{*)},
showstringspaces=false,
morestring=[b]",
morestring=[d],
tabsize=3,
extendedchars=false,
sensitive=true,
breaklines=false,
basicstyle=\small,
captionpos=b,
columns=[l]flexible,
identifierstyle={\ttfamily\color{black}},
keywordstyle=[1]{\ttfamily\color{violet}},
keywordstyle=[2]{\ttfamily\color{green}},
keywordstyle=[3]{\ttfamily\color{blue}},
keywordstyle=[4]{\ttfamily\color{blue}},
keywordstyle=[5]{\ttfamily\color{red}},
stringstyle=\ttfamily,
commentstyle={\ttfamily\color{green}},
literate=
    {\\forall}{{\color{green}{$\forall\;$}}}1
    {\\exists}{{$\exists\;$}}1
    {<-}{{$\leftarrow\;$}}1
    {=>}{{$\Rightarrow\;$}}1
    {==}{{\code{==}\;}}1
    {==>}{{\code{==>}\;}}1
    {->}{{$\rightarrow\;$}}1
    {<->}{{$\leftrightarrow\;$}}1
    {<==}{{$\leq\;$}}1
    {\#}{{$^\star$}}1 
    {\\o}{{$\circ\;$}}1 
    {\@}{{$\cdot$}}1 
    {\/\\}{{$\wedge\;$}}1
    {\\\/}{{$\vee\;$}}1
    {++}{{\code{++}}}1
    {~}{{\ }}1
    {\@\@}{{$@$}}1
    {\\mapsto}{{$\mapsto\;$}}1
    {\\hline}{{\rule{\linewidth}{0.5pt}}}1
}[keywords,comments,strings]

\lstnewenvironment{coq}{\lstset{language=Coq}}{}
\newif\ifprintcomments

\newcommand{\lstl}[1]{\lstinline[language=Coq, basicstyle=\normalsize]{#1}}

\newcommand{\sref}[1]{\S\ref{sec:#1}}
\newcommand{\flabel}[1]{\label{fig:#1}}
\newcommand{\fref}[1]{Figure~\ref{fig:#1}}

\newcommand{\ie}{\emph{i.e.}, }
\newcommand{\eg}{\emph{e.g.}, }

\newcommand{\Var}[1]{\textsf{\textit{#1}}}

\newcommand{\T}[1]{\textsf{\textbf{#1}}}

\newcommand{\clitp}[1]{\texttt{\normalsize lit\_pat #1}}
\newcommand{\cholep}[0]{\texttt{\normalsize hole\_pat}}
\newcommand{\cinholep}[2]{\texttt{\normalsize inhole\_pat #1 #2}}
\newcommand{\cinholept}[0]{\texttt{\normalsize inhole\_pat}}
\newcommand{\cnilp}[0]{\texttt{\normalsize nil\_pat\_c}}
\newcommand{\cconsp}[2]{\texttt{\normalsize cons\_pat\_c #1 #2}}

\newcommand{\cnamep}[2]{\texttt{\normalsize name\_pat #1 #2}}
\newcommand{\cnamept}[0]{\texttt{\normalsize name\_pat}}
\newcommand{\cntp}[1]{\texttt{\normalsize nt\_pat #1}}
\newcommand{\cntpt}[0]{\texttt{\normalsize nt\_pat}}
\newcommand{\inholep}[2]{\ensuremath{\T{in-hole} \; #1 \; #2}}
\newcommand{\inholet}[0]{\T{in-hole}}
\newcommand{\namep}[2]{\ensuremath{\T{name}\; #1\; #2}}
\newcommand{\namet}{\T{name}}
\newcommand{\ntp}[1]{\ensuremath{\T{nt}\; \Var{#1}}}

\newcommand{\const}[0]{\T{cons}}
\newcommand{\consp}[2]{\ensuremath{\const\; #1 \; #2}}

\newcommand{\produ}[2]{\ensuremath{(#1, #2)}}

\newcommand{\lit}[1]{\ensuremath{\T{#1}}}
\newcommand{\nil}[0]{\ensuremath{\T{nil}}}

\newcommand{\cconst}[2]{\texttt{\normalsize cons\_term\_c #1 #2}}
\newcommand{\cconstt}[0]{\texttt{\normalsize cons\_term\_c}}
\newcommand{\chdcontt}[0]{\texttt{\normalsize hd\_contxt}}
\newcommand{\chdcont}[2]{\texttt{\normalsize hd\_contxt #1 #2}}
\newcommand{\ctailcontt}[0]{\texttt{\normalsize tail\_contxt}}
\newcommand{\ctailcont}[2]{\texttt{\normalsize tail\_contxt #1 #2}}
\newcommand{\cholet}[0]{\texttt{\normalsize hole\_contxt\_c}}
\newcommand{\clitt}[1]{\texttt{\normalsize lit\_term #1}}
\newcommand{\cnilt}[0]{\texttt{\normalsize nil\_term\_c}}

\newcommand{\holet}[0]{\T{hole}}
\newcommand{\plug}[2]{#1[\![ #2 ]\!]}
\newcommand{\mtch}[5]{\ensuremath{#1 \vdash #2 : {#3}_{#4} \; | \; #5}}
\newcommand{\decom}[7]{\ensuremath{#1 \vdash #2 = \plug{#3}{#4} : {#5}_{#6} \; | \; #7}}
\newcommand{\bunion}[2]{\ensuremath{#1 \sqcup #2}}

\newcommand{\delprod}[3]{\ensuremath{#1 \setminus \produ{#2}{#3}}}
\newcommand{\ccon}[2]{\ensuremath{#1 ++ \; #2}} 

\newcommand{\mevbody}[0]{\ensuremath{\mathsf{M_{ev\_gen}}}}
\newcommand{\mev}[0]{\ensuremath{\mathsf{M_{ev}}}}

\newcommand{\selectev}[6]{\ensuremath{\mathsf{select} (#1, #2, #3, #4, #5, #6)}}
\newcommand{\combine}[5]{\ensuremath{\mathsf{combine} \; (#1, #2, #3, #4, #5)}}
\newcommand{\named}[2]{\ensuremath{\mathsf{named} (#1, #2)}}

\newcommand{\mtchpair}[3]{\ensuremath{(#2, #3)_{#1}}}
\newcommand{\empdec}[1]{\ensuremath{\bullet_{#1}}}
\newcommand{\nempdecev}[4]{\ensuremath{(#2, #3)^{#4}_{#1}}}

\newcommand{\subts}{\ensuremath{<_{\mathsf{subt}}}}
\newcommand{\subt}[2]{\ensuremath{#1 \; \subts \; #2}}

\newcommand{\subterms}[3]{\ensuremath{\mathsf{subterms} \; #1 \; #2 \; #3}}
\newcommand{\mtup}[3]{\ensuremath{(#1, #2, #3)}}
\newcommand{\inpconsord}[2]{\ensuremath{#1 \inpconsordsym #2}}
\newcommand{\inpconsordsym}{\ensuremath{<_{\mathtt{t}}}}
\newcommand{\nictup}[2]{\ensuremath{(#1, #2)}}
\newcommand{\ninpconsordsym}{\ensuremath{<_{\mathtt{p \times g}}}}
\newcommand{\ninpconsord}[2]{\ensuremath{#1 \ninpconsordsym #2}}
\newcommand{\lexprodsym}[1]{\ensuremath{<^{#1}_{\mathtt{t \times p \times g}}}}
\newcommand{\lexprod}[3]{\ensuremath{#2 \lexprodsym{#1} #3}}

\newcommand{\col}[0]{CoLoR}
\newcommand{\redexk}[0]{RedexK}
\usepackage{paralist}

\usepackage{hyperref}

\usepackage{xspace}

\usepackage{tikz-cd}

\usepackage[justification=centering]{caption} 

\usepackage[toc,page]{appendix}

\begin{document}

\title{Redex $\rightarrow$ Coq: towards a theory of decidability of Redex's reduction semantics}


\author{Mallku Soldevila \and
        Rodrigo Ribeiro \and
        Beta Ziliani}

\institute{M.~Soldevila \at
  FAMAF, UNC and CONICET (Argentina) \\
  \email{mes0107@famaf.unc.edu.ar}\\
  ORCID: 0000-0002-8653-8084
\and 
  B.~Ziliani \at
  FAMAF, UNC and Manas.Tech (Argentina)\\
  \email{beta@mpi-sws.org}\\
  ORCID: 0000-0001-7071-6010
\and
  R.~Ribeiro \at
  DECOM, UFOP (Brazil)\\
  \email{rodrigo.ribeiro@ufop.edu.br}\\
  ORCID: 0000-0003-0131-5154
}

\begin{abstract}
  We propose the first steps in the development of a tool to automate the
  translation of Redex models into a (hopefully) semantically equivalent model in Coq,
  and to provide tactics to help in the certification of fundamental properties of such models.
  The work is heavily based on a model of Redex's semantics developed by 
  Klein et al. By means of a simple generalization of the matching problem 
  in Redex, we obtain an algorithm suitable for its mechanization in Coq, 
  for which we prove its soundness properties and its correspondence with 
  the original solution proposed by Klein et al. In the process, we
  also adequate some parts of our mechanization to better prepare it for
  the future inclusion of Redex features absent in the present model, like
  its Kleene-star operator. Finally, we discuss future avenues of development 
  that are enabled by this work.


  

  
\end{abstract}

\maketitle

\section{Introduction}
\label{sec:introduction}
\paragraph{Motivation}
Redex \citep{plt} is a DSL, built on top of the Racket programming language, 
that allows for the mechanization of reduction semantics models and formal 
systems. It also includes a variety of tools 
for testing them: unit testing, random generators of terms for random testing of 
properties, stepper for step-by-step reduction sequences. It has been successfully 
used for the mechanization of large semantics models of real programming languages 
(\eg JavaScript \cite{guh, get}; Python \cite{python}; Scheme \cite{schemeop}; and
Lua 5.2 \cite{ dls, lua-ppdp20, lua_auto}); the development of
tools for program analysis (like, again, \cite{lua-ppdp20}, to check for a
particular kind of \emph{well-behavedness} of Lua 5.2 programs; 
\cite{modular_type_sound}, for checking type-soundness of syntactic language
extensions that introduces high-level programming concepts). Other, particular
uses cases, involve the mechanization of operational semantics for virtual machines
specialised for running reactive programs \cite{reactive}, or even mechanizing a model 
of Redex itself, as is done in \cite{plt}.

Redex's approach to semantics engineering involves a philosophy about documents
that specify semantics models, which can be summarized as
``semantic models as software artifacts'' \citep{runredex}, and a lightweight
development of models that focuses on a quick transition between specification
of models and testing of their properties. These virtues of Redex enable it as
a useful tool with which to perform the first steps of a formalization effort. 
Nonetheless, when a given model seems to be thoroughly tested and mature, it
could be of use to actually prove its desired properties, since no amount of
testing can guarantee the absence of errors.

As an example, consider the experience with \cite{guh}, a major step into the development
of semantics models for JavaScript, dubbed $\lambda_{JS}$. It has been reported 
that, even after the mechanisation of their semantics model with Redex, and after intensive 
testing of an interpreter derived from the mechanization against major
implementations of JavaScript, other researchers found a missing case in the
semantics.\footnote{See \url{https://blog.brownplt.org/2012/06/04/lambdajs-coq.html}}
The missing case caused certain terms to get \emph{stuck}, breaking a progress 
property claimed for the model. This called for a revision of the model, in 
search for any other flaw, but equipped with a proof assistant. To this end, the
researchers mechanized $\lambda_{JS}$ entirely into Coq.

At the moment there is no other way to tackle such task: the model must be
written again entirely into a proof assistant. Besides being a time-consuming
process, another downside is that the translation into the proof assistant
may be guided just by an intuitive understanding of the behavior of the
mechanization in Redex. Intuitive understanding that could differ from the actual 
behavior of the model in Redex. This is so, since the tool implements a particular
meaning of reduction semantics with evaluation contexts, offering an expressive language
to the user that includes several features, useful to express concepts like context-dependent
syntactic rules. The actual semantics of this language may not coincide with 
what the researcher understands (see \cite{semcontext} for a development of this
issue).

Our proposal, to assist in mitigating the described situation, consists in helping 
the user with the automatic translation of a given model in Redex, into an equivalent model 
in Coq. The interpretation, of the resulting model in Coq, will be done through a shallow 
embedding in this proof assistant of Redex's actual semantics. In that regard, we note that there 
already exist several implementations of some of the concepts of reduction 
semantics with evaluation contexts (see \sref{related_work} for a detailed description
of the available options). However, some features of Redex, like its support for evaluation 
contexts and its semantics for a Kleene's closure of patterns, are particular to the tool. 
To gain trust about the correspondence between the original model in Redex and its transpiled
version into Coq, it may be preferable to have a direct explanation of this last model
in terms of Redex's own behavior, avoiding codifying Redex's concepts on top of another 
model of reduction semantics.

\paragraph{Summary of the contributions.}
In this work we present a first step into the development of a tool to  
automate the translation of a Redex model into a semantically equivalent model 
in Coq, and to provide automation to the proof of essential properties of such
models. The present work is heavily based on the model of Redex's semantics
developed by Klein et al. \citep{semcontext} (from now on, \redexk). In summary:
\begin{itemize}
  \item We mechanize \redexk{} in Coq. In the process, we develop a proof of termination
        for the matching algorithm, which enables its
        mechanization into Coq as a regular primitive recursion.

  \item We modify \redexk{} to prepare it for the future
        addition of features, like Redex's Kleene closure of patterns and the development
        of tactics to decide about properties of reduction semantics models.

  \item We prove soundness properties of the matching algorithm with respect to its 
        specification.

  \item We prove the correspondence of our algorithm with respect to the original
        proposal present in \redexk.
\end{itemize}

The reader is invited to download the accompanying source code from 
\url{github.com/Mallku2/redex-into-coq}.

The remainder of this paper is structured as follows:
\sref{red_semantics} presents a brief introduction to reduction semantics,
as presented in Redex; \sref{redex_coq} offers a general overview of our
mechanization in Coq; \sref{soundness} presents the main soundness results
proved within our mechanization; \sref{related_work} discuss about related
work from the literature of the area; finally, \sref{conclusion} summarizes
the results presented in this paper and discusses future venues of research
enabled by this first iteration of our tool.


\section{Redex}
\label{sec:red_semantics}
In this section, we will present a brief introduction to Redex's main concepts, limiting our attention
to the concepts that are relevant to our tool in this first iteration of the 
development.

As a running example, we show how to mechanize a fragment of $\lambda$-calculus with normal order 
reduction, in Redex. For a better introduction to these topics, the reader can consult \citet{plt, runredex} 
and the original paper on which our mechanization is based~\citep{semcontext}.
Also, its reference manual presents the most up-to-date information about Redex's 
features.



Redex can be viewed as a particular implementation of the semantics of reduction semantics 
with evaluation contexts (RS). Reduction semantics~\citep{plt} follows the intellectual 
tradition of providing a theory about the concepts that are expressed by a given 
language, just in terms of relations over terms of said language. 
This tradition is embodied in theories like $\lambda$-calculus, proposed by Alonzo 
Church~\citep{barendregt} as a way to explain and study functions, in terms
of rewriting relations. While its capabilities to express computations
were already known~\citep{barendregt}, it was rediscovered as a way to formally describe
programming languages later, by researchers like Peter J. Landin (\eg see \citet{mech_eval, calgol}). 
These syntactic theories ended up 
being useful to \emph{explain} several different phenomena and mechanisms present in programming languages, 
in a concise and abstract way: \eg evaluation strategies, parameter passing style in function-calls, complex 
control-flow features and state~\citep{syntheo}.\footnote{Note the emphasis put in the word ``explain'': 
not every researcher would concur with the idea that a relation over terms is actually explaining said
terms (for example, see \citet{stoy}, page 9).}

In the context of semantics for programming languages, 
terms represent actual programs (and, maybe, some semantics elements), and the relations over 
programs can represent dynamic and static semantics relations. In order to define these structures 
that contain terms (languages and relations), the user of Redex uses a language of \emph{patterns}.
These patterns constitute a highly expressive language, whose semantics is explained 
specifying which terms match against a given pattern. This formalization of the notion
of matching against Redex's patterns is the main focus of \citet{semcontext}, together 
with the development of an algorithmic interpretation of this specification. Mechanize this work
in Coq, solving problems like finding a primitive recursive implementation of the matching process, 
constitutes the main work presented in this paper.

As a simple introductory example, consider \fref{redex_grammar}, where it is shown
the definition of a grammar that captures terms of a call-by-value $\lambda$-calculus, where we impose 
normal-order evaluation, using \emph{evaluation contexts}. The grammar contains non-terminals
\lstl{e} (representing any $\lambda$ term), \lstl{v} (representing values; in this case, only
$\lambda$-abstractions), \lstl{x} (representing variables) and \lstl{E} (representing
evaluation contexts, to be explained below). The right-hand-side of the productions
of each non-terminal are shown on the right of the \lstl{::=} symbol. 

\begin{figure}
  \begin{center}
  \includegraphics[scale=0.4]{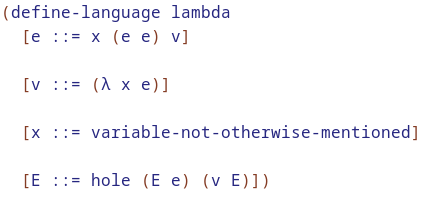}
  \end{center}
  \caption{Definition of a language in Redex.}
  \flabel{redex_grammar}
\end{figure}

Productions of non-terminals \lstl{e} and \lstl{x} are standard. In the case of non-terminal
\lstl{x}, the right-hand-side of its only production is defined with a pattern (\lstl{variable-not-otherwise-mentioned})
that has a context-sensitive meaning: the terms that \emph{match against} non-terminal \lstl{x}
(\ie the terms that can be produced by \lstl{x}), are only those that do not match against
the remaining non-terminals (\ie the terms that cannot be produced by the remaining non-terminals).

Within the toolbox of RS, the syntactic notion of \emph{contexts} is a useful device to express 
context-sensitive rules and concise definitions. A \emph{context} is a term with a special position 
(or positions) denoted with a marker, a \emph{hole}. RS then offers ways to refer to these contexts, to reason over them and 
to manipulate them, through the operations of \emph{decomposition} of a given term into a 
context and another sub-term, and \emph{plugging} a given term into the hole of a given context.
Decomposition is referred through a special pattern, that expresses the way into
which a given term must be decomposed. Plugging is its dual concept, and it is
denoted in a similar way, typically being the position where it occurs on a given 
definition what distinguishes it from a decomposition.

Context themselves may represent the literal context where a given term appears within a program. 
\emph{Evaluation contexts} are a special category of contexts, used typically in programming languages' semantics,
that point into a single 
position within a program, indicating where we should look for the next redex during a reduction.
If we are interested in a deterministic dynamic semantics, we could use evaluation contexts
to impose a particular reduction (or evaluation) order: for a well-defined notion of evaluation contexts,
it should be possible to decompose every program into a unique evaluation context and a sub-term,
that should be a redex (according to the given dynamic semantics).

Returning to \fref{redex_grammar}, the productions of non-terminal \lstl{E} indicate that
an evaluation context could be a single hole, or a context of the form $\Var{E'} \; \Var{e}$,
where \Var{E'} is another evaluation context; or a context of the form $\Var{v} \; \Var{E'}$.
Note that the consequence of this definition is that, for a given $\lambda$ term of the
form $\Var{e}_1 \; \Var{e}_2$, we will evaluate it in a normal-order fashion: that is, the
next redex should be looked into $\Var{e}_2$ only if $\Var{e}_1$ is already a value; 
otherwise, we start looking for the redex within $\Var{e}_1$.

\begin{figure}
  \begin{center}
  \includegraphics[scale=0.4]{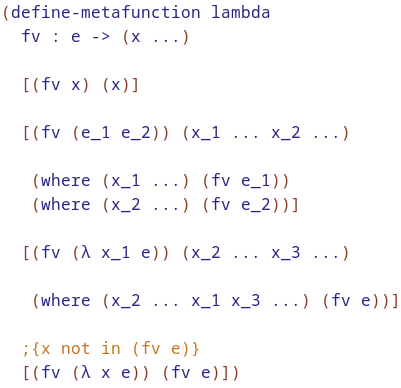}
  \end{center}
  \caption{Definition of a meta-function in Redex: \emph{free variable} in $\lambda$ terms.}
  \flabel{redex_fv}
\end{figure}

We can also define a notion of \emph{free variable} in $\lambda$ terms with a \emph{meta-function} \lstl{fv}, whose equations are listed in \fref{redex_fv}.
Note that we can define the (run-time checked) signature of the function,
\lstl{fv : e -> (x ...)}, which explains that \lstl{fv} receives a $\lambda$ term,
and returns a list of 0 or more variables (pattern \lstl{x ...}, to be explained below).
After the signature, we have 4 equations explaining which are the free variable in: a
term that is a single variable \lstl{x}; an application $\mathtt{e_1} \; \mathtt{e_2}$;
a $\lambda$ abstraction whose formal parameter ($\mathtt{x_1}$) occurs free in
its body (\lstl{e}); and a $\lambda$ abstraction whose formal parameter (\lstl{x}) does not
occur free in its body (\lstl{e}). Note that the second and third equation contain
side-conditions, in the form of a clause \lstl{where}: their semantics dictate that
these conditions hold if the expression to the right, matches against the pattern on 
the left. For example, the first \lstl{where} clause of the second equation holds
if the expression $\mathtt{(fv \; e_1)}$ (an evaluation of \lstl{fv} over term $\mathtt{e_1}$),
 matches against the pattern $\mathtt{(x_1 \; ...)}$, to be explained below.

The definition of \lstl{fv} shows a feature of Redex which is particular to it: the 
\emph{Kleene closure} of a given pattern, which serves to express the idea of
``zero or more terms'' that match against a given pattern. It is denoted as a
pattern followed by \emph{...} (\ie a mathematical ellipsis). In the previous
figure, it was used to define the domain of \lstl{fv} (a list of ``0
or more variables'', with pattern \lstl{(x ...)}), and in the second and third
equation, within the \lstl{where} clauses and when expressing the final 
value of \lstl{fv}. For example, as mentioned previously, the first \lstl{where}
clause of the second equation imposes a condition that holds only when
the expression $\mathtt{fv \; e_1}$ matches against the pattern $\mathtt{x_1 \; ...}$,
meaning that $\mathtt{fv \; e_1}$ must evaluate to a list of 0 or more
variables. Redex bind that list with $\mathtt{x_1 \; ...}$, and we can use this pattern
whenever we want to refer to this list. In particular, the value of
\lstl{fv} over the abstraction of this second equation, means that we return the variables to which
$\mathtt{fv \; e_1}$ evaluated ($\mathtt{x_1 \; ...}$), followed by the 
variables to which $\mathtt{fv \; e_2}$ evaluated ($\mathtt{x_2 \; ...}$):
that is, $\mathtt{x_1 \; ... \; x_2 \; ...}$. Note that, in the
\lstl{where} clause of the fourth equation, we are asking for $\mathtt{fv \; e}$
to match against the pattern $\mathtt{x_2 \; ... \; x_1 \; x_3 \; ...}$, where
$\mathtt{x_1}$ is the formal parameter and \lstl{e} is the body of the $\lambda$ 
abstraction whose free variables we want to obtain. This means that we are
forcing the situation where the formal parameter appears in the list of free
variables of \lstl{e}. In other words, the third equation refers to the
case where the formal parameter of the $\lambda$ abstraction appears free in
its body.

The interesting aspect of the previous language of patterns is that it allows us to 
enforce context-dependent restrictions, through many devices. For example, by 
repeating sub-patterns, within a given pattern, the user can enforce the repetition 
of elements into a given list of terms or any other part of a phrase. 
For example, the pattern $(\mathtt{x\_1} \; \mathtt{x\_1})$ only matches against a list 
of 2 equal variables. Also, we can force some sub-terms to be different from
the rest, by using the suffix $\_!\_$ after each pattern whose matching term we 
want to distinguish from the rest. For example, the pattern 
$(\mathtt{x\_!\_} \; \mathtt{x\_!\_})$ only matches against a list of 2 different
variables. These devices, to enforce context-dependent rules, can be exploited 
to define languages, but also any relations over their terms.
 
\begin{figure}
  \begin{center}
  \includegraphics[scale=0.4]{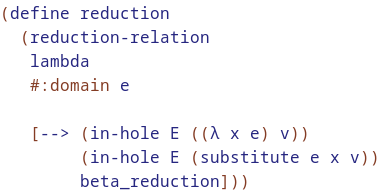}
  \end{center}
   \caption{Definition of a semantics relation in Redex.}
  \flabel{redex_sem}
\end{figure}

Finally, \fref{redex_sem} depicts the definition of the compatible closure (with
respect to evaluation contexts \lstl{E}) of a call-by-value $\beta$-contraction,
in Redex (note the keyword \lstl{reduction-relation}). The single reduction rule 
shown explain 2 things: how $\beta$-contractions are done, using a generic
capture-avoiding substitution function (\lstl{substitute}); and the order in which
those contraction can occur, effectively imposing the order of evaluation described
by contexts \lstl{E}. The rule states that, if a given term can be decomposed 
between some context \lstl{E} and some abstraction application $\mathtt{((\lambda \; x \; e) \; v)}$
(condition expressed through the pattern $\mathtt{(in-hole\; E\; ((\lambda \; x \; e) \; v))}$), 
then, the original term reduces to the phrase resulting from plugging \lstl{(substitute e x v)} 
(\ie capture-avoiding substitution of the formal parameter \lstl{x}, by the value \lstl{v}, into 
the abstractions' body \lstl{e}) into the context \lstl{E} (which is expressed through the 
pattern $\mathtt{in-hole \; E \; (substitute \; e \; x \; v)}$). Finally, the resulting 
relation will be the \emph{least} relation that satisfies the given conditions. That is, 
these definitions can be translated as the usual Coq's inductive relations.

For reasons of space, and to keep our example simple, we eluded the definition 
of the capture-avoiding substitution function. This can be defined as our previous specification 
of function \lstl{fv} (plus some escaping to Racket code). However, Redex itself provides 
a general mechanism to get a substitution function \emph{by free}, requiring from us only 
to indicate the bounding occurrences of variables in the constructions of our language, 
and their scope. This feature is not included in \redexk{}, and neither is it considered in
our mechanization.

The previous brief introduction to Redex served the purpose of introducing some
features with which we will be dealing when working with \redexk{}. We avoid features
that are not covered in said model. Also, not every capability previously described 
is covered in \redexk{}, though we need to mention them in order to easily implement
our model of $\lambda$-calculus: we are talking about the Kleene closure of patterns, 
used when defining meta-function \lstl{fv} (\fref{redex_fv}); and the pattern 
\lstl{variable-not-otherwise-mentioned}, used to define $\lambda$ variables (\fref{redex_grammar}).

\section{Expressing Redex in Coq.}
\label{sec:redex_coq}
In this section, we introduce the main ideas behind our implementation in Coq.
Later, in \sref{soundness}, we will provide a specification of the obtained
algorithm, proofs asserting the correspondence between the algorithm and its
specification, and between our specification and the one provided in the
original work.

To introduce the simpler parts of the mechanization, we will show listings of our 
source code together with some natural language explanation. The more complex portions of
the mechanization (like the matching/decomposition algorithm), will be described 
more abstractly. In that way, while being faithful to our mechanization, we will 
avoid the expected complexities of an actual implementation with a 
dependently-typed language like Coq.

\subsection{Language of terms and patterns}
We begin the presentation by introducing our mechanized version of the language of 
terms and patterns. We ask for some reasonable decidability properties 
about the language that we use to describe a given reduction semantics model. These 
standard properties will be useful to develop our mechanization in its present version,
but also in the prospective future of the development.


\subsubsection{Symbols}
\label{sec:symbols}
We require for the elements of the language of terms and patterns (literals, non-terminals 
and sub-indexes used in the patterns) to be equipped with a decidable definitional equality.
To formalize these properties we take advantage of \lstl{stdpp}'s~\citep{stdpp}
typeclass \lstl{EqDecision}. We abstract all of these assumptions into the module type \lstl{Symbols}, 
shown in \fref{symbols}.

In order to implement an instantiation of a module of type \lstl{Symbols},
we ask the user for the type of literals, \emph{pattern variables} (or the sub-indexes 
of patterns mentioned in \sref{red_semantics}) and non-terminals of the grammar: 
computational types \lstl{lit}, \lstl{var} and
\lstl{nonterm}, respectively. We also ask for proofs showing that these types
are also instances of typeclass \lstl{EqDecision}.
Naturally, we do not want to burden the user with these proofs. A requirement for
our future transpiler from Redex to Coq should be that it must be able to automatically
build these proofs, something that is feasible within Coq.
\begin{figure}

\begin{lstlisting}
Module Type Symbols.
  (* literals for both, pats and terms *)
  Parameter lit: Set.
  (* names in name_pat *)
  Parameter var: Set.
  (* representation of the non-terms of a given grammar, for patterns nt *)
  Parameter nonterm: Set.

  (* some assumptions to ease the reasoning about decidability *)
  Parameter nonterm_eq_dec : EqDecision nonterm.
  Parameter var_eq_dec : EqDecision var.
  Parameter lit_eq_dec : EqDecision lit.
End Symbols.
\end{lstlisting}
\caption{A module type capturing assumptions about several atomic elements of the grammar.}
\label{fig:symbols}
\end{figure}

\subsubsection{Terms}

\begin{figure}
\begin{lstlisting}
Inductive term : Set :=
  | lit_term     : lit -> term
  | list_term_c : list_term -> term
  | contxt_term : contxt -> term

  with list_term : Set :=
  | nil_term_c   : list_term
  | cons_term_c : term -> list_term -> list_term

  with contxt : Set :=
  | hole_contxt_c : contxt
  | list_contxt_c : list_contxt -> contxt

  (* hd_contxt and tail_contxt point into a position of a list of terms  *)
  with list_contxt : Set :=
  | hd_contxt    : contxt -> list_term -> list_contxt
  | tail_contxt : term -> list_contxt -> list_contxt.
\end{lstlisting}
\caption{Language of terms.}
\label{fig:terms}
\end{figure}

In the original paper,
terms are classified according to: their structure, or, if they act as a context or not. 
According to their structure, terms are classified as atomic literals or with a 
binary-tree structure. In our case, we will generalize the notion of ``terms with
structure''. One of the most prominent features absent in \redexk{} is Redex's
Kleene closure of patterns. Such patterns match against (or describe) lists of
0 or more terms. In order to be able to
include this feature in a future iteration of our model, we begin by generalizing
the notion of structured terms. We will allow them to be lists of 0 or more terms.
Non-empty lists can also be considered as binary trees, but where the right sub-tree
of a given node is always a list. We will enforce that shape through types.

The language of terms is presented in \ref{fig:terms}. Terms as literals are built
with constructor \lstl{lit_term}, while structured terms are captured and enforced
through a type, \lstl{list_term}. Structured terms can be an empty list,
built with \lstl{nil_term_c}, or a list with one term as its head, a some
list as its tail, using constructor \lstl{cons_term_c}. Finally, we define an 
injection into terms, \lstl{list_term_c}.

The other kind of terms considered in \redexk{} are contexts. 
Being a context involves not only the existence of a hole marking some position into a term (as
mentioned in \sref{red_semantics}). It also involves including information
describing where to find that marked position, in the context itself, to help the 
algorithms of decomposition and plugging. That information consists in a path
from the root of the term (seen as a tree) to the leaf that contains the hole. 
To that end, the authors of \redexk{} defined a notion of context that, if it is not 
just a single hole, it contains a \emph{tag} indicating where to look for the hole: 
either into the left or the right sub-tree of the context.
We preserve the same idea, adapted to our presentation of structured terms: now,
a hole could mark the head or the tail of a list, and we add to the contexts' tags
indicating that information.

We introduce the type \lstl{contxt}, to represent and enforce through types the
notion of contexts. These contexts can be just a single hole, built with constructor
\lstl{hole_contxt_c}, or a list of terms with some position marked with a hole.
In order to guarantee the presence of a hole into this last kind of contexts, we
introduce the type \lstl{list_contxt}. These contexts can point into the 
first position of a given list, constructed with \lstl{hd_contxt}, or the tail,
constructed with \lstl{tail_contxt}. Finally, we have injections from 
\lstl{list_contxt} into \lstl{contxt} (\lstl{list_contxt_c}), and
from \lstl{contxt} into \lstl{term} (\lstl{contxt_term}). These 
injections, naturally, are used later as coercions.

\subsubsection{Patterns} 
\label{sec:patterns}

\begin{figure}
\begin{lstlisting}
  Inductive pat : Set :=
  | lit_pat    : lit -> pat
  | hole_pat   : pat
  | list_pat_c : list_pat -> pat
  | name_pat   : var -> pat -> pat
  | nt_pat     : nonterm -> pat
  | inhole_pat : pat -> pat -> pat

  with list_pat : Set :=
  | nil_pat_c  : list_pat
  | cons_pat_c : pat -> list_pat -> list_pat.
\end{lstlisting}
\caption{Language of patterns.}
\label{fig:patterns}
\end{figure}

As mentioned in \sref{red_semantics}, Redex offers a language of patterns 
with enough expressive power to state even context-dependent restrictions. We
mechanize the same language of patterns as presented in \redexk{}, with the 
required change to accommodate our generalization done to structured terms,
as explained in the previous sub-section. The language of patterns is presented
in \ref{fig:patterns}.

Pattern \clitp{\Var{l}} matches only against a single literal 
\Var{l}. Pattern \cholep{} matches against a context that is
just a single hole. In order to describe the new category of structured terms
that we presented in the previous subsection, we add a new category of patterns
enforced through type \lstl{list_pat}. From this category of patterns, 
pattern \cnilp{} matches against a list of 0 terms, while pattern 
\cconsp{\Var{p}$_{hd}$}{\Var{p}$_{tl}$} matches against a list of terms, whose first term 
matches against pattern \Var{p}$_{hd}$, and whose tail matches against the pattern 
\Var{p}$_{tl}$. Finally, we have a injection from this category of patterns into
the type \lstl{pat}: \lstl{list_pat_c}.

Context-dependent restrictions are imposed through pattern \cnamep{\Var{x}}{\Var{p}}.
This pattern matches against a term \Var{t} that, in turn, must match against 
pattern \Var{p}. As a result, the pattern \cnamep{\Var{x}}{\Var{p}} introduces
a context-dependent restriction in the form of a \emph{binding}, that assigns 
\emph{pattern variable} \Var{x} to term \Var{t}. Data-structures to keep 
track of this information will be introduced later, but for the moment, just consider that 
during matching some structures are used to keep track of all of this context-dependent 
restrictions that have the form of a binding between a pattern variable and a term.
If, at the moment of introducing the binding to \Var{x}, there exist another
binding for the same variable but with respect to a term different than \Var{t},
the whole matching fails. This could happen if, for example, pattern 
\cnamep{\Var{x}}{\Var{p}} is just a sub-pattern from another pattern, and there is already
a sub-pattern of the form \cnamep{\Var{x}}{\Var{p'}}, where \Var{p'} already
matched against a term different than \Var{t}.

With the language of patterns we can describe the grammar of our language, as well
as specify the many kinds of relations over terms of our language. For example, the
following pattern (taken from the grammar of the $\lambda$-calculus shown in 
\fref{redex_grammar}):
\begin{quote}
\cconsp{(\cntp{\Var{e}})}{(\cconsp{(\cntp{\Var{e}})}{\cnilp})}\footnote{For simplicity, we
avoid mentioning the injection of this value into type \lstl{pat}, through \lstl{list_pat_c}.}
\end{quote}
would represent the right-hand-side of the production that indicates that a $\lambda$
term applied to another $\lambda$ term, is a valid term.
As seen, patterns themselves can contain mentions to non-terminals of our grammar: pattern 
\cntp{\Var{e}} matches against a term \Var{t}, if there exist a
production from non-terminal \Var{e}, whose right-hand-side is a pattern 
\Var{p} that matches against term \Var{t}.

Finally, pattern \cinholep{\Var{p}$_c$}{\Var{p}$_h$} matches against some term \Var{t}, if 
\Var{t} can be decomposed between some context \Var{C}, that matches
against pattern \Var{p}$_c$, and some term \Var{t'}, that matches against
pattern \Var{p}$_t$. It should be possible to plug \Var{t'} into context 
\Var{C}, recovering the original term \Var{t}. Note that the information 
contained in the tag of each kind of non-empty context, that indicates where 
to find the hole, helps in this process: at each step the process looks, 
either, into the head of the context or into its tail.


\subsubsection{Decidability of predicates about terms and patterns}

We want to put particular emphasis on the development of tools to recognize the 
decidability of predicates about terms and patterns. This could serve as a good
foundation for the future development of tactics to help the user automate
as much as possible the process of proving arbitrary statements about the
user's reduction semantics models.

As a natural consequence of our first assumptions about the atomic elements of
the languages of terms and patterns, presented in \sref{symbols}, we can 
also prove decidability results about definitional equalities among terms
and patterns. Another straightforward consequence involves the decidability
of definitional equalities between values of the many
data-structures involved in the process of matching. Future efforts will be 
put in developing further this minimal theory about decidability. See \sref{conclusion}.

\subsubsection{Grammars}
The notion of grammar in Redex, as presented in \sref{red_semantics}, is
modeled in \redexk{} as a finite mapping between non-terminals and sets of
patterns. Our intention is not to force some particular representation for grammars.
As a first step, we axiomatize some assumptions about grammars through a module type. 
We begin by defining a production of the grammar, simply, as a pair inhabiting \lstl{nonterm * pat}, and
we define a \lstl{productions} type as a list of type production. We also ask for the existence of computational type 
\lstl{grammar}, a constructor for grammars 
(\lstl{new_grammar : productions -> grammar}), the possibility of testing
\emph{membership} of a production with respect to a grammar, and to be possible to 
\emph{remove} a production from a grammar (\lstl{remove_prod}).\footnote{That is, we ask
for the possibility of building a new grammar from a given one, that does not 
contain some particular production of the later grammar.} We ask for some
notion of \emph{length} of grammars, and that \lstl{remove_prod} actually affects
that length in the expected way. This will be useful to guarantee the
termination property of the matching algorithm (see \sref{well_founded}).
Finally, we ask for some reasonable decidability properties for these types and 
operations: decidability of definitional equalities among values of the previous 
types, and, naturally, for the testing of membership of a production with respect to a 
given grammar.

Abstracting these previous types and properties in a module type (\lstl{Grammar}), could 
serve in the future when developing further our theory of decidability for 
the notion of RS implemented in Redex. As a simple example, separating the type 
\lstl{productions} from the actual definition of the type \lstl{grammar}, allows
for the encapsulation of properties in the type \lstl{grammar} itself, that 
specifies something about the inhabitants of \lstl{productions}. Some decidability 
results depend on a grammar whose productions are restricted
in some particular way.\footnote{For example, while the general language intersection 
problem for context-free grammars (CFG) is non-decidable, the intersection between
a regular CFG and a non-recursive CFG happens to be decidable 
\cite{cfg_intersection}.}

For this first iteration, we provide an instantiation of the previous module
type with a grammar implemented using a list of productions. Here, the type
\lstl{grammar} does not impose new properties over the type \lstl{productions}.
We also provide a minimal theory to reason about \emph{grammars as lists}, that
helps in proving the required termination and soundness properties of the matching 
algorithm. This is required since our previous axiomatization of grammars, through 
module type \lstl{Grammar}, is not strong enough to prove every desired property of our 
algorithm. A goal for a next iteration would be to take advantage of the experience with this
development, and strengthen our axiomatization of grammars.

\subsubsection{Remaining data-structures}
\label{sec:data_structs}
We end this sub-section with a brief description of the most important 
remaining data-structures, needed to implement matching and decomposition:
\begin{itemize}
  \item \lstl{binding : var * term}: a representation of a context-dependent 
        restriction, introduced by the pattern \cnamept{}, as described in 
        \sref{patterns}.

   \item \lstl{decom_ev : term -> Set}: a dependently-typed representation
         of a decomposition of a given term \Var{t}, between a context and a 
         sub-term. We make this type dependent on \Var{t}, and include in 
         \lstl{decom_ev} some evidence of soundness of the decomposition.
   \item \lstl{mtch_ev : term -> Set}: a dependently-typed representation
         of one result from a matching/decomposition of a given term \Var{t},
         against some pattern.
         It contains an instance of \lstl{binding}, and an instance of
         \lstl{decom_ev} depending on \Var{t} itself.
\end{itemize}

Their actual purpose will be clear in \sref{match_decom_alg}, when introducing the 
matching/decomposition algorithm. Also, functions to manipulate values of the
previous types will be presented as needed.

\subsection{Matching and decomposition}
The first challenge that we encounter when trying to mechanize \redexk{}, is 
that of finding a primitive recursive algorithm to express matching and 
decomposition. The original algorithm from \redexk{} is not a primitive 
recursion, for reasons that will be clear below. However, the theory developed in 
the paper, to check the soundness of this algorithm and to characterize the inputs 
over which it actually converges to a result, helped us to recapture the matching 
and decomposition process as a \emph{well-founded recursion}.

\subsubsection{Well-founded relation over the domain of matching/decomposition}
\label{sec:well_founded}
In Coq, a well-founded recursion is presented as a primitive recursion over the 
evidence of \emph{accessibility} of a given element (from the domain of the 
well-founded recursion), with respect to a given \emph{well-founded relation} $R$. 
That is, it is a 
primitive recursion over the proof of a statement that asserts that, from a given 
actual parameter $x$ over which we are evaluating a function call, there 
is only a finite quantity of elements which are \emph{smaller} than $x$, according to 
relation $R$. These smaller elements are the ones over which recursive 
function calls can be evaluated. In other words: $R$ does not 
contain infinite decreasing chains, and, hence, the number of recursive function 
calls is always finite. Such relation $R$ is called well-founded.

The actual steps of matching/decomposition will be presented in detail below. 
But, for the moment, in pursuing a well-founded recursive definition for 
the matching/decomposition process, let us observe that, for a given grammar 
$\Var{G}$, pattern $\Var{p}$ and term $\Var{t}$, the matching/decomposition of 
$\Var{t}$ against $\Var{p}$ involves, either:

\begin{enumerate}
\item \label{first} Steps where the input term \Var{t} is \emph{decomposed} or
      \emph{consumed}.\\

\item Steps where there is no input consumption, but, either:\\
  \begin{enumerate}
    \item \label{second_pat} The pattern \Var{p} is decomposed or consumed.\\
    \item \label{second_prod} The productions of the grammar \Var{G} are considered,
      searching for a suitable pattern against which the matching should proceed.
   \end{enumerate}
\end{enumerate}

Step \ref{first} corresponds, for example, to the case where \Var{t} is a list 
of terms of the form \cconst{\Var{t}$_{hd}$}{\Var{t}$_{tl}$}, and \Var{p} is a list of patterns of the
form \cconsp{\Var{p}$_{hd}$}{\Var{p}$_{tl}$}. Here, the root of each tree (\Var{t} and \Var{p}) match,
and the next step involves checking if \Var{hd} matches against pattern 
\Var{hd'}, and if \Var{tl} matches against \Var{tl'}. In each case, some
part of \Var{t} has been consumed, and the following steps involve considering
for matching some proper sub-term of \Var{t}. Clearly, we can perform only a
finite amount of these kind of steps.

Step \ref{second_pat} corresponds, for example, to the case where pattern \Var{p}
has the form \cnamep{\Var{x}}{\Var{p'}}: as described in \sref{patterns}, the next step
in matching/decomposition involves checking if pattern \Var{p'} matches 
against \Var{t}. Here, the step does not involve consumption of input term
\Var{t}, but it does involve a recursive call to matching/decomposition over
a proper sub-pattern of \Var{p}. Again, we can perform only a finite amount 
of these kind of steps.

Finally, step \ref{second_prod} corresponds to the case of pattern \cntp{\Var{n}}, 
which implies looking for productions of \Var{n} in \Var{G} that match against
\Var{t}. Here, there is no reduction of terms and this process does not 
neccesarily imply the reduction of patterns.

If not because for the pattern \cntpt, it could be easily argued that the process 
previously described is indeed an algorithm. Now, if we do take into account 
\cntpt{} patterns, termination in the general case does no longer holds. 
In particular, non-termination could be observed with a grammar 
\Var{G} \emph{left-recursive} and a given non-terminal \Var{n} that witnesses the 
left-recursion of \Var{G}. Matching against pattern \cntp{\Var{n}}, following the
described process, could get stuck repeating the step of searching into the 
productions of \Var{n}, without any consumption of input: from pattern 
\cntp{\Var{n}} we could reach to the same pattern \cntp{\Var{n}}, over and over again.

Indeed, the described matching algorithm does not deal with left-recursion, as
is argued in \cite{semcontext}. There, the property of left-recursion
is captured by providing a relation $\rightarrow_{G}$ that order patterns as 
they appear during the previously described phase of the matching process, when 
the input term is not being consumed, but there is decomposition of a pattern and/or
searching into the grammar, looking for a proper production to continue the
matching. Then, a left-recursive grammar would be one that makes the chains of the
previous relation to contain a repeated pattern. In particular, during matching,
we could begin with a pattern \cntp{\Var{n}} and reach the same pattern without consuming 
input, repeating this process over and over again. We mention here said definition:
\begin{definition}[Left-recursion \cite{semcontext}]
\label{def:left_recursion}
A grammar \Var{G} is left recursive if \Var{p} $\rightarrow^+_{\Var{G}}$ \Var{p} 
for some pattern \Var{p}, where $\rightarrow^+_{\Var{G}}$ is the transitive 
(but not reflexive) closure of 
$\rightarrow_{\Var{G}}$ : \lstl{pat} $\times$ \lstl{pat}, the least relation 
satisfying the following conditions:
\begin{itemize}
\item[] \cntp{\Var{n}} $\rightarrow_{\Var{G}}$ \Var{p}, if \Var{p} $\in$ 
\Var{G}(\Var{n}) 

\item[] \cnamep{\Var{x}}{\Var{p}} $\rightarrow_{\Var{G}}$ \lstl{p}

\item[] \cinholep{\Var{p}$_c$}{\Var{p}$_h$} $\rightarrow_{\Var{G}}$ \Var{p}$_c$

\item[] \cinholep{\Var{p}$_c$}{\Var{p}$_h$} $\rightarrow_{\Var{G}}$ \Var{p}$_h$,
if pattern \Var{p}$_c$ matches against \cholet{}
\end{itemize}
\end{definition}

Then, if, for a non left-recursive grammar \Var{G} and non-terminal \Var{n} from
\Var{G}, it is the case that \Var{p} $\not \rightarrow^+_{\Var{G}}$ \Var{p} for 
any pattern \Var{p}, it must be the case that also
\mbox{\cntp{\Var{n}} $\not \rightarrow^+_{\Var{G}}$ \cntp{\Var{n}}}.
This means that, when searching for productions of \Var{n} in \Var{G}, and as 
long as the matching/decomposition is in the stage captured by $\rightarrow_{\Var{G}}$, 
(\ie no consumption of input), it should be possible to \emph{discard} the productions 
from \Var{G} being tested.

The previous observation helps us argue that, provided that \Var{G} is non 
left-recursive, when the matching process enters the stage of non-consumption of 
input, this phase will eventually finalize: either, the pattern under consideration is 
totally decomposed and/or we run out of productions from \Var{G}. In what follows, 
we will assume \emph{only} non-left-recursive grammars. This will not
impose a limitation over our model of Redex, since it only allows such kind of grammars.

We will exploit the previous observations to build a well-founded relation
over the domain of our matching/decomposition function. The technique that we will
use will consist in, first, modeling each phase in isolation through a particular relation. 
There will be a relation 
$\inpconsordsym : $ \lstl{term} $ \rightarrow $ \lstl{term} $\rightarrow$ \lstl{Prop} 
explaining what happens to the input when it is being consumed, and a relation 
$\ninpconsordsym$ : \lstl{pat} $\times $ \lstl{grammar} $\rightarrow $ \lstl{pat} $\times $ \lstl{grammar} $\rightarrow$ \lstl{Prop},
explaining what happens to the 
pattern and the grammar when there is no consumption of input. We will also 
prove the well-foundedness of each relation. The final well-founded relation for the 
matching/decomposition function will be the \emph{lexicographic product} 
of the previous relations, a well-known method to build new well-founded relations out of 
other such relations \citep{lexico}. We will parameterize this relation by
the original grammar, to be able to recover the original productions when
needed.\footnote{We will present in \sref{match_spec_sec} the situations where this is needed.} 
For a given grammar \Var{g}, we will denote this last relation with 
$\lexprodsym{\Var{g}}$. Note that its type is: 
\begin{center}
  \lstl{term} $\times $ \lstl{pat} $\times $ \lstl{grammar} $\rightarrow $ 
                          \lstl{term} $\times $ \lstl{pat} $\times $ \lstl{grammar} $\rightarrow$ \lstl{Prop} 
\end{center}

For a tuple $(\Var{t}, \Var{p}, \Var{G})$ to be 
related with another \emph{smaller} tuple $(\Var{t'}, \Var{p'}, \Var{G'})$, according 
to $\lexprodsym{\Var{g}}$, it must happen the following:

\begin{center}
$\inpconsord{\Var{t'}}{\Var{t}} \vee 
(\Var{t'} = \Var{t} \wedge \ninpconsord{(\Var{p'}, \Var{G'})}{(\Var{p}, \Var{G})})$
\end{center}

This expresses the situations where there is actual progress in the 
matching/decomposition algorithm towards a result: either there is consumption of 
input or the phase of production searching and decomposition of the pattern 
progresses towards its completion. Note that, however, this definition shows 
that the lexicographic product is a more general relation, that contains chains 
of tuples that do not necessarily model what happens during matching and 
decomposition: if $\inpconsord{\Var{t'}}{\Var{t}}$, then
$\lexprod{\Var{g}}{(\Var{t'}, \Var{p'}, \Var{G'})}{(\Var{t}, \Var{p}, \Var{G})}$, 
for some grammar \Var{g},
regardless of what $(\Var{p'}, \Var{G'})$ and $(\Var{p}, \Var{G})$ actually
are. Later, when presenting the relations that form this lexicographic product,
we will also specify which are the actual chains that we will consider when
modeling the process of matching and decomposition. We will refer to these 
last kind of chains as the \emph{interesting chains} or \emph{chains of interest}.

The previous means that we will define a more general relation, that is simpler to
define and to work with, but that still retains the desired properties: it will
be well-founded and will contain the chains of interest, besides other meaningless
chains.

For our implementation, we will simply use Coq's standard library implementation
of lexicographic product on pairs:
\begin{lstlisting}
slexprod : forall A B : Type, (A -> A -> Prop) -> (B -> B -> Prop) -> A * B -> A * B -> Prop
\end{lstlisting}
That is, for a given grammar \Var{g}, $\lexprodsym{\Var{g}}$ will be defined in terms
of \lstl{slexprod}. As noted, this relation is well-founded provided we are able to prove the 
well-foundedness of its composing relations. Hence, the following type is inhabited:
\begin{lstlisting}
forall (A B : Type) (leA : A -> A -> Prop) (leB : B -> B -> Prop),
well_founded leA -> well_founded leB -> well_founded (slexprod A B leA leB)
\end{lstlisting}

Note that \lstl{well_founded} \Var{le} simply codifies the type stating that the relation \Var{le} is well-founded:
\begin{lstlisting}
well_founded = fun (A : Type) (R : A -> A -> Prop), forall a : A, Acc R a
\end{lstlisting}

Where, for a given relation \lstl{R} and element \lstl{a} in its domain, 
\lstl{Acc R a} is the type of proofs showing that \lstl{a} is \emph{accesible} for
relation \lstl{R}: informally, there is only a finite amount of elements smaller than
\lstl{a}, according to \lstl{R} (see \cite{cpdt}, section 7.1, for a more detailed 
presentation of the concept).

In what follows, we will present our definition for the relations $\inpconsord{}$ and
$\ninpconsord$. Fortunately, they describe simple processes for which it is possible
to prove their well-foundedness without resorting to complex arguments.

\subsubsection{Input consumption}
\label{sec:input_consumption_sec}

As stated in the previous sub-section, $\inpconsordsym$ should model how
the input term is \emph{consumed} or \emph{decomposed} during
matching and decomposition. This amounts to relate a term \Var{t'} with
another term \Var{t}, in that order, if from term \Var{t} we can reach
term \Var{t'} during a recursive evaluation of matching/decomposition
of \Var{t} against some pattern. A reasonable definition for this relation
can be, simply, this: $\inpconsordsym = \subts$, where $\subts$ denotes the
relation:
\begin{center}
  \lstl{subterm_rel} : \lstl{term} $\rightarrow$ \lstl{term} $\rightarrow$ \lstl{Prop}
\end{center}
that links a term with each of its sub-terms.\footnote{In turn, \lstl{subterm_rel}
is defined for verification purposes of the matching/decomposition algorithm.
See \sref{match_decom_alg}.}
That is, 
$\inpconsord{\Var{t'}}{\Var{t}}$ if \Var{t'} is just any sub-term of \Var{t}. 
For the actual specification of
matching and decomposition, this definition is enough (as we will see
when introducing the process in \sref{match_decom_alg}). This does not avoid
for more exotic patterns, that could be introduced in the future, to have 
a different behavior on input consumption, in such a way that some 
recursive evaluation involves a term that is not an actual sub-term of
the original input term.\footnote{Consider, for example, context-dependent
rules such as in the pattern 
$(\mathtt{x\_!\_} \; \mathtt{x\_!\_})$, described in \sref{red_semantics}.
It only matches against a list of 2 different variables: its semantics
cannot be explained by considering only recursive evaluations of matching
between proper sub-patterns and sub-terms of the input term.}
 For this first iteration of our tool, we just 
acknowledge that this could happen in a future version of the language of 
patterns of Redex. Hence, we will assume the existence of a relation 
$\inpconsordsym$, with the purpose already described, and that, for the 
time being, it is exactly $\subt{}{}$.

While in this sub-section we are concerned with $\inpconsordsym$, there
is still a related issue that also involves our parameterized lexicographic
product of relations: recall that, given a grammar \Var{g}, 
$\lexprod{\Var{g}}{(\Var{t'}, \Var{p'}, \Var{G'})}{(\Var{t}, \Var{p}, \Var{G})}$
holds if and only if:

\begin{center}
$\inpconsord{\Var{t'}}{\Var{t}} \vee 
(\Var{t'} = \Var{t} \wedge \ninpconsord{(\Var{p'}, \Var{G'})}{(\Var{p}, \Var{G})})$
\end{center}

If what it actually holds is $\inpconsord{\Var{t'}}{\Var{t}}$, then the pair
$(\Var{p'}, \Var{G'})$ can be anything. As we mentioned in the previous sub-section,
this means that $\lexprodsym{\Var{g}}$ contains chains that do not necessarily 
model an actual process of matching/decomposition. In our case, the chains
that will be of interest are the ones where, when there is actual input
consumption (\ie $\inpconsord{\Var{t'}}{\Var{t}}$), then \Var{p'} is some
sub-pattern of \Var{p} (following rules to be introduced in \sref{match_spec_sec}) and 
$\Var{G'} = \Var{g}$. That is, after a step of input consumption, we \emph{re-install}
the original grammar \Var{g} in the tuple. This is needed since we need to
guarantee that, if a pattern of the form \cntp{\Var{n}}, for some non-terminal 
\lstl{n}, appears during matching/decomposition, we have at our disposal
every production of \lstl{n}, for production searching. The only situation where
it is guaranteed that we do not have to worry about this situation, is \emph{after}
the appearance of pattern \cntp{\Var{n}}, and before the next step where input consumption
occurs. This is the phase \ref{second_prod} mentioned in \sref{well_founded}:
when a pattern like \cntp{\Var{n}} appears, the process of production searching
begins. And because \Var{g} is non-left recursive (recall that we only assume
such kind of grammars; see \sref{well_founded}), it is guaranteed
that we will not need to look for another production of \lstl{n}, as long
as this phase of the matching/decomposition process continues. 
This will become more clear when introducing the actual algorithm of 
matching/decomposition and its specification, beginning in \sref{match_spec_sec}.

\subsubsection{Pattern and production consumption}
\begin{figure}
  \begin{mathpar}
    \inference{}
    {\ninpconsord
      {(\Var{p}_c,\Var{G})}
      {(\cinholep{\Var{p}$_c$}{\Var{p}$_h$}, \Var{G})}}

    \inference{}
    {\ninpconsord
      {\nictup{\Var{p}_h}{\Var{G}}}
      {\nictup{\cinholep{\Var{p}$_c$}{\Var{p}$_h$}}{\Var{G}}}}

    \inference{}
    {\ninpconsord
      {\nictup{\Var{p}}{\Var{G}}} 
      {\nictup{\cnamep{\Var{x}}{\Var{p}}}{\Var{G}}}}\\

    \inference
    {\Var{p} \in \Var{G}(\Var{n})\\ 
      \Var{G'} = \delprod{\Var{G}}{\Var{n}}{\Var{p}}}
    {\ninpconsord
      {\nictup{\Var{p}}{\Var{G'}}} 
      {\nictup{\cntp{\Var{n}}}{\Var{G}}}}    
\end{mathpar}
\caption{Consumption of pattern and productions.}
\flabel{no_input_cons}
\end{figure}
We now turn to the specification of $\ninpconsordsym$, which explains how 
evolve the pattern and the grammar (over which we interpret the non-terminals
from the pattern), when there is no input consumption. This stage of the
matching/decomposition algorithm corresponds to phases \ref{second_pat} and
\ref{second_prod} described in \sref{well_founded}. Recall that, in this case,
the algorithm entered a phase where the pattern is being decomposed or
productions from some non-terminal are being tested, to see if 
matching/decomposition can continue. Note that the evolution of the pattern 
in this stage of the algorithm is already described in Definition \ref{def:left_recursion},
in \sref{well_founded}. Indeed, $\ninpconsordsym$ will be defined just
considering the inverse of the relation showed in said definition, plus some 
particular considerations about grammars.

We present in \fref{no_input_cons} the definition of $\ninpconsordsym$. 
Matching a term \Var{t} against a pattern of the form \cinholep{\Var{p}$_c$}{\Var{p}$_h$},
means trying to decompose the term between some context that matches against 
pattern \Var{pc}, and some sub-term of \Var{t} that matches against pattern 
\Var{p}$_h$. In doing so, the first step involves a decomposition process (to
be specified later in \sref{decom_spec_sec}), that begins working over the whole term \Var{t}, 
and with respect to just the sub-pattern \Var{p}$_c$. Hence, this step does not
involve input consumption, but it does involve considering a reduced pattern:
\Var{pc}. We just capture this simple fact through $\ninpconsordsym$, by stating
that $\ninpconsord{(\Var{p}_c,\Var{G})}{(\cinholep{\Var{p}$_c$}{\Var{p}$_h$}, \Var{G})}$
holds, for any grammar \Var{G}. Note that we preserve the grammar.

In the particular case that \Var{p}$_c$ matches against \cholet{}, then there is
no actual decomposition of the term \Var{t}. This means that, when looking
for said sub-term of \Var{t} that matches against pattern \Var{p}$_h$, we will
still being considering the whole input term \Var{t}: no input consumption occurred
as a result of extracting out from \Var{t} a context that matches against \Var{p}$_c$. 
Again, we just capture this simple fact 
by stating that $\ninpconsord{(\Var{p}_h,\Var{G})}{(\cinholep{\Var{p}$_c$}{\Var{p}$_h$}, \Var{G})}$
holds, for any grammar \Var{G}. Note that we also preserve the grammar, and
that we do not force this situation to hold only when the pattern \Var{pc}
matches against \cholet{}. This results in a relation $\ninpconsordsym$
that contains some chains of tuples that do not correspond to the matching/decomposition
algorithm. This is not a problem, since it also contain the chains that we
need, and the resulting definition is simpler.

The case for the pattern \cnamep{\Var{x}}{\Var{p}} can be explained on the same
basis as with the previous cases: matching term \Var{t} against pattern
\cnamep{\Var{x}}{\Var{p}} involves, first, trying to match the whole term
\Var{t} against the sub-pattern \Var{p}. There is no input consumption involved
 in this first step, but there is a reduction of the pattern. We also preserve
the grammar in this step.


Finally, the last case refers to the pattern \cntp{\Var{n}}: it involves considering
each production of non-terminal \Var{n} in \Var{G}. Here it is assumed that \Var{G} 
contains the correct set of productions that remain to be tested (an invariant property
about \Var{G} through our algorithm, to be justified below). Then, we continue the process considering a grammar \Var{G'}
that contains every production from \Var{G}, except for \produ{\Var{n}}{\Var{p}}: the 
already considered production of non-terminal \Var{n} with right-hand-side \Var{p}. We
denote it stating that \Var{G'} equals the expression $\delprod{\Var{G}}{\Var{n}}{\Var{p}}$.

The previously mentioned invariant about \Var{G} will be
maintained through the chains of interest of our lexicographic product $\lexprodsym{\Var{g}}$,
for a given grammar \Var{g} over which we begin the matching/decomposition
process. Preserving this invariant involves maintaining unaltered the grammar
over which we interpret the non-terminals of the pattern (as shown in the first 
3 cases of \fref{no_input_cons}) in the absence of pattern decomposition, 
but allowing $\ninpconsordsym$ to consider
a \emph{smaller} grammar once some production is tested (last case in
\fref{no_input_cons}), on the basis of the non-left-recursivity of the grammar
being considered. Finally, preserving the invariant also involves 
reestablishing to the original grammar \Var{g}, once the matching returns to 
input consumption (something to be specified \sref{match_spec_sec}).

A final concern about $\ninpconsordsym$ is related with convincing ourselves that this
relation does not contain infinite decreasing chains: \ie that it is well-founded.
Looking again at \fref{no_input_cons} we observe that, at each step, either the 
pattern is being reduced, or the grammar considered contains less productions.
Hence, for example, a simple proof by a nested induction, first, on the size of the 
grammar and, at each case, structural induction on the pattern, suffices to show
the well-foundedness of $\ninpconsordsym$.

\subsubsection{Specification of matching}
\label{sec:match_spec_sec}
We now turn to the task of modifying the original specification for matching and
decomposition from \redexk. As we will see, our specification defines a simple
generalization of the original problem, as presented in \cite{semcontext}: here,
we will allow for the matching and decomposition algorithm to interpret the
non-terminals in the pattern by looking for productions from some arbitrary
grammar, not just the original grammar, during some specific phase of the process.

The specification for \redexk{} consists of 2 mutually inductive formal systems,
that help to build proofs for judgments that speak about matching and decomposition.
We will begin by presenting the formal system that specifies the notion of
matching. Judgments about matching have the form $\mtch{\Var{G}}{\Var{t}}{\Var{p}}{}{\Var{b}}$, 
stating that pattern \Var{t} matches against pattern \Var{p}, under the productions from
grammar \Var{G}, producing the bindings \Var{b} (which could be an empty set of bindings,
denoted with $\oslash$). The non-terminals that may appear
on pattern \Var{p} will be interpreted in terms of the productions from \Var{G}.
Hence, the formal system that allows us to build proofs for such judgments, explains
the semantics of matching against a given pattern.

\begin{figure}
\begin{mathpar}
\inference{}
          {\mtch
           {\Var{G}}
           {\clitt{\Var{a}}}
           {(\clitp{\Var{a}})}
           {\Var{G'}}
           {\oslash}}

\inference{}
          {\mtch
           {\Var{G}}
           {\cholet}
           {\cholep}
           {\Var{G'}}
           {\oslash}}

\inference{}
          {\mtch
              {\Var{G}}
              {\cnilt}
              {\cnilp}
              {\Var{G'}}
              {\oslash}}

\inference{\mtch
           {\Var{G}}
           {\Var{t}}
           {\Var{p}}
           {\Var{G'}}
           {\Var{b}}}
         {\mtch
           {\Var{G}}
           {\Var{t}}
           {(\cnamep{\Var{x}}{\Var{p}})}
           {\Var{G'}}
           {\bunion{\Var{b}}{\{ (\Var{x}, \Var{t}) \}}}}

\inference{\Var{p} \in \Var{G'}(\Var{n})\\
           \mtch
           {\Var{G}}
           {\Var{t}}
           {\Var{p}}
           {\Var{\delprod{\Var{G'}}{\Var{n}}{\Var{p}}}}
           {\Var{b}}}
         {\mtch
           {\Var{G}}
           {\Var{t}}
           {(\cntp{\Var{n}})}
           {\Var{G'}}
           {\oslash}}\\
         
\inference{\mtch
           {\Var{G}}
           {\Var{t}_{hd}}
           {(\Var{p}_{hd})}
           {\Var{G}}
           {\Var{b}_{hd}}\\
           \mtch
           {\Var{G}}
           {\Var{t}_{tl}}
           {(\Var{p}_{tl})}
           {\Var{G}}
           {\Var{b}_{tl}}}
         {\mtch
           {\Var{G}}
           {\cconst{\Var{t}$_{hd}$}{\Var{t}$_{tl}$}}
           {(\cconsp{\Var{p}$_{hd}$}{\Var{p}$_{tl}$})}
           {\Var{G'}}
           {\bunion{\Var{b}_{hd}}{\Var{b}_{tl}}}
           }

\inference{\decom
                {\Var{G}}
                {\Var{t}}
                {\Var{C}}
                {\Var{t}_h}
                {(\Var{p}_c)}
                {\Var{G'}}
                {\Var{b}_c}\\
                \subt{\Var{t}_h}{\Var{t}}\\
          \mtch
           {\Var{G}}
           {\Var{t}_h}
           {(\Var{p}_h)}
           {\Var{G}}
           {\Var{b}_h}}
         {\mtch
           {\Var{G}}
           {\Var{t}}
           {(\cinholep{\Var{p}$_c$}{\Var{p}$_h$})}
           {\Var{G'}}
           {\bunion{\Var{b}_c}{\Var{b}_h}}}

\inference{\decom
                {\Var{G}}
                {\Var{t}}
                {\cholet}
                {\Var{t}}
                {(\Var{p}_c)}
                {\Var{G'}}
                {\Var{b}_c}\\
           \mtch
            {\Var{G}}
            {\Var{t}}
            {(\Var{p}_h)}
            {\Var{G'}}
            {\Var{b}_h}}
          {\mtch
             {\Var{G}}
             {\Var{t}}
             {(\cinholep{\Var{p}$_c$}{\Var{p}$_h$})}
             {\Var{G'}}
             {\bunion{\Var{b}_c}{\Var{b}_h}}}
       \end{mathpar}
\caption{Modified specification of matching.}
\flabel{match_spec}
\end{figure}

Here, we will consider a generalization of this problem: our formal system will
serve to build proofs for judgments of the form $\mtch{\Var{G}}{\Var{t}}{\Var{p}}{\Var{G'}}{\Var{b}}$,
stating almost the same as the previous formal system, with the particular difference
that, \emph{initially}, we interpret the non-terminals from \Var{p} looking for their 
productions in some arbitrary grammar \Var{G'} (that is what the notation $\Var{p}_{\Var{G'}}$
tries to suggest). Only when input consumption 
begins, we will turn to the original grammar \Var{G}. \fref{match_spec} presents a
simplified fragment of our formal system. Following a top-down, left-to-right order,
the first rule states that a term of the form \clitt{\Var{a}} (a literal) only matches 
against a pattern of the form \clitp{\Var{a}}, producing no bindings. Here, the grammars
play no role. The second rule and third rules can be understood on the same basis.

The fourth rule explains the way in which a pattern of the form $\mbox{\cnamep{\Var{x}}{\Var{p}}}$
introduces context-dependent restrictions, when a given term \Var{t} successfully
matches against it. This implies that sub-pattern \Var{p} matches against \Var{t}, 
producing bindings \Var{b}, and a new binding $(\Var{x}, \Var{t})$ can be added to \Var{b}. 
This is done through the disjoint-union of bindings, denoted with 
$\bunion{\Var{b}}{\{ (\Var{x}, \Var{t}) \}}$. This operation is defined only if there 
is no binding for \Var{x} in \Var{b}, or, if $\Var{b}(\Var{x}) = \Var{t}$. Note that, 
given that we recursively prove matching for the whole input term \Var{t} (\ie no input
consumption occurred), we preserve the grammar \Var{G'}. That is, we are following the chains from the well-founded
relation $\ninpconsordsym$ (\fref{no_input_cons}). This semantics accounts for the 
behavior shown in \sref{red_semantics}, when referring to the sub-terms of a given term, 
after the matching, through the names presented in the pattern. See, for example, 
\fref{redex_fv}, where the pattern being described is used in defining the 
equations that capture the meta-function \lstl{fv} from the $\lambda$-calculus.

The fifth rule explains what it means for a term \Var{t} to match against a pattern
\cntp{\Var{n}}, when the non-terminals of this pattern (in this case, just \Var{n})
are \emph{initially} interpreted in terms of the productions of some arbitrary grammar \Var{G'}:
then, that matching is successful if there exist some $\Var{p} \in \Var{G'}(\Var{n})$,
such that \Var{t} matches against \Var{p}, when its non-terminals are \emph{initially}
interpreted under the productions from the grammar $\Var{\delprod{\Var{G'}}{\Var{n}}{\Var{p}}}$.
Recall that this means that this last grammar will be used as long as there is no input 
consumption, or there is no other appearance of a pattern \cntpt{}. Again, we are
following the chains from $\ninpconsordsym$. Also, the non-left-recursivity of the grammars
being considered guarantee that this replacement of the grammars is semantics-preserving:
we will not need another production from \Var{n}, as long as there is no input consumption.
Finally, note that this match does not produce bindings 

The sixth rule describes matching of a term that represents a list of terms
(\cconst{\Var{t}$_{hd}$}{\Var{t}$_{tl}$}) against a pattern that also describes a list
of terms, through a list of patterns (\cconsp{\Var{p}$_{hd}$}{\Var{p}$_{tl}$}). We consider 
this matching partially successful if the head of the list of terms, \Var{t}$_{hd}$, matches against 
the head of the list of patterns, \Var{p}$_{hd}$, producing some bindings \Var{b}$_{hd}$.
Note that, given that this last match is done over an actual sub-term of the
original input, we \emph{re-install} the original grammar \Var{G}, to interpret the
non-terminals from \Var{p}$_{hd}$. We also ask for the tail of the input list of terms,
\Var{t}$_{tl}$, to match against the tail of the list of patterns, \Var{p}$_{tl}$.
Again, this match is done over a sub-term of the input term, hence, we consider
the original grammar \Var{G}. The non-left-recursivity of the grammars being considered
does not interfere with the possibility of both, \Var{p}$_{hd}$ and \Var{p}$_{tl}$,
include patterns \cntpt{}.

If successful, the previous match produces some bindings
\Var{b}$_{tl}$. Finally, it will be possible to prove the match between the original list of
terms and patterns, if the disjoint union between \Var{b}$_{hd}$ and \Var{b}$_{tl}$ 
is defined. Consider what would happen if \Var{p}$_{hd}$ and \Var{p}$_{tl}$
contain \cnamept{} patterns that introduce contradictory restrictions: the match
should fail. This also shows how these context-dependent restrictions operate, to
impose conditions over different parts of a given term. Finally, the cases for 
contexts \chdcontt{} and \ctailcontt{} (contexts in the form of list of terms, with
one special hole), are totally analogous to this case.

The last 2 cases in \fref{match_spec} refer to the matching of a term 
\Var{t} against a pattern of the form \cinholep{\Var{p}$_c$}{\Var{p}$_h$}. This
operation is successful when we can decompose term \Var{t} between some context,
that matches against pattern \Var{p}$_c$, and some sub-term, that matches against
pattern \Var{p}$_h$. In order to fully formalize what this matching means, we
need to explain what \emph{decomposition} means. \redexk{} specifies this notion
through another formal system, whose adaptation to our work we present in 
the following sub-section. The original system allows us to build proofs for judgments of
the form $\decom{\Var{G}}{\Var{t}}{\Var{C}}{\Var{t'}}{\Var{p}}{}{\Var{b}}$,
meaning that we can decompose term \Var{t}, between some context \Var{C}, that
matches against pattern \Var{p}, and some sub-term \Var{t'}. The decomposition
produces bindings \Var{b}, and the non-terminals from pattern \Var{p} are
interpreted through the productions present in grammar \Var{G}. In our case,
we modify this judgments (and the formal system itself), by generalizing them
in the same way done for the matching judgments: now, we consider judgments
of the form $\decom{\Var{G}}{\Var{t}}{\Var{C}}{\Var{t'}}{\Var{p}}{\Var{G'}}{\Var{b}}$,
meaning almost the same as the previous decomposition judgment, with the possibility
of interpreting the non-terminals in \Var{p}, initially, through the productions
from some arbitrary grammar \Var{G'} (that is, $\Var{p}_{\Var{G'}}$). We will explain 
in detail this formal system in the next sub-section.

Returning to the cases about \cinholept{} patterns, in \fref{match_spec}, note
that we distinguish the case where the decomposition step actually consumes some
portion from \Var{t}, from the case where it does not. The first situation (described
in the first rule for \cinholept{}) means that  context \Var{C} is not simply a 
hole, and $\Var{t}_h$ is an actual proper sub-term of \Var{t}: \ie $\subt{\Var{t}_h}{\Var{t}}$.
Also, note that the decomposition is proved interpreting (initially) the non-terminals 
from $\Var{p}_c$ with production from the arbitrary grammar \Var{G'} ($(\Var{p}_c)_{\Var{G'}}$).
And the proof of the matching between $\Var{t}_h$ and $\Var{p}_h$ is done temporally
interpreting the non-terminals of this last pattern with productions from the original
grammar \Var{G} ($(\Var{p}_c)_{\Var{G}}$).

The second rule for \cinholept{} considers the possibility that the initial
decomposition did not consume some part of the input term \Var{t}. That is,
\Var{p}$_{c}$ matched against a single hole (\cholep{}). In that case, the
decomposition did not produce an actual sub-term of \Var{t}, and the following
match against pattern \Var{p}$_{h}$ is done with the whole input term. Hence,
$(\Var{p}_{h})_{\Var{G'}}$. Note that, again, in both cases of \cinholept{}, 
the final set of bindings in the judgment is the result of the disjoint 
union of bindings from the decomposition of the term, and the matching
with its sub-term.

\subsubsection{Specification of decomposition}
\label{sec:decom_spec_sec}
The final part of the specification concerns specifically with the process of
decomposition. That is, part of the semantics of the \cinholept{} pattern.
As already mentioned, the original specification of this concept comes in the
form of a formal system to prove judgments of the form 
$\decom{\Var{G}}{\Var{t}}{\Var{C}}{\Var{t'}}{\Var{p}}{}{\Var{b}}$ (explained 
previously), that we generalize to judgments of the form\\
$\decom{\Var{G}}{\Var{t}}{\Var{C}}{\Var{t'}}{\Var{p}}{\Var{G'}}{\Var{b}}$, that
we also introduced in the previous sub-section. \fref{decomp_spec} presents
a simplified fragment of the modified formal system. We describe the rules
following a top-down order.

\begin{figure}
\begin{mathpar}         
\inference{}
          {\decom
                {\Var{G}}
                {\Var{t}}
                {\cholet}
                {\Var{t}}
                {\cholep}
                {\Var{G'}}
                {\oslash}}\\

\inference{\decom
                {\Var{G}}
                {\Var{t}_{hd}}
                {\Var{C}}
                {\Var{t'}_{hd}}
                {(\Var{p}_{hd})}
                {\Var{G}}
                {\Var{b}_{hd}}\\
           \mtch
                {\Var{G}}
                {\Var{t}_{tl}}
                {(\Var{p}_{tl})}
                {\Var{G}}
                {\Var{b}_{tl}}}
         {\decom
                {\Var{G}}
                {\cconst{\Var{t}$_{hd}$}{\Var{t}$_{tl}$}}
                {(\chdcont{\Var{C}}{\Var{t}$_{tl}$})}
                {\Var{t'}_{hd}}
                {(\cconsp{\Var{p}$_{hd}$}{\Var{p}$_{tl}$})}
                {\Var{G'}}
                {\bunion{\Var{b}_{hd}}{\Var{b}_{tl}}}
           }\\

\inference{\mtch
                {\Var{G}}
                {\Var{t}_{hd}}
                {(\Var{p}_{hd})}
                {\Var{G}}
                {\Var{b}_{hd}}\\
             \decom
                {\Var{G}}
                {\Var{t}_{tl}}
                {\Var{C}}
                {\Var{t'}_{tl}}
                {(\Var{p}_{tl})}
                {\Var{G}}
                {\Var{b}_{tl}}}
         {\decom
                {\Var{G}}
                {\cconst{\Var{t}$_{hd}$}{\Var{t}$_{tl}$}}
                {(\ctailcont{\Var{t}$_{hd}$}{\Var{C}})}
                {\Var{t'}_{tl}}
                {(\cconsp{\Var{p}$_{hd}$}{\Var{p}$_{tl}$})}
                {\Var{G'}}
                {\bunion{\Var{b}_{hd}}{\Var{b}_{tl}}}
           }\\

\inference{\Var{p} \in \Var{G'}(\Var{n})\\
           \decom
                {\Var{G}}
                {\Var{t}}
                {\Var{C}}
                {\Var{t'}}
                {\Var{p}}
                {\delprod{\Var{G'}}{\Var{n}}{\Var{p}}}
                {\Var{b}}}
          {\decom
                {\Var{G}}
                {\Var{t}}
                {\Var{C}}
                {\Var{t'}}
                {(\cntp{\Var{n}})}
                {\Var{G'}}
                {\oslash}}\\

\inference{\decom
                {\Var{G}}
                {\Var{t}}
                {\Var{C}_c}
                {\Var{t}_c}
                {(\Var{p}_c)}
                {\Var{G'}}
                {\Var{b}_{c}}\\
           \subt{\Var{t}_c}{\Var{t}}\\
           \decom
                {\Var{G}}
                {\Var{t}_c}
                {\Var{C}_h}
                {\Var{t}_h}
                {(\Var{p}_h)}
                {\Var{G}}
                {\Var{b}_{h}}}
        {\decom
          {\Var{G}}
          {\Var{t}}
          {(\ccon{\Var{C}_c}{\Var{C}_h})}
          {\Var{t}_h}
          {(\cinholep{\Var{p}$_c$}{\Var{p}$_h$})}
          {\Var{G'}}
          {\bunion{\Var{b}_{c}}{\Var{b}_{h}}}}\\

\inference{\decom
                {\Var{G}}
                {\Var{t}}
                {\cholet}
                {\Var{t}}
                {(\Var{p}_c)}
                {\Var{G'}}
                {\Var{b}_{c}}\\
           \decom
                {\Var{G}}
                {\Var{t}}
                {\Var{C}_h}
                {\Var{t}_h}
                {(\Var{p}_h)}
                {\Var{G'}}
                {\Var{b}_{h}}}
        {\decom
          {\Var{G}}
          {\Var{t}}
          {(\ccon{\cholet}{\Var{C}_h})}
          {\Var{t}_c}
          {(\cinholep{\Var{p}$_c$}{\Var{p}$_h$})}
          {\Var{G'}}
          {\bunion{\Var{b}_{c}}{\Var{b}_{h}}}}\\

\inference{\decom
              {\Var{G}}
              {\Var{t}}
              {\Var{C}}
              {\Var{t'}}
              {\Var{p}}
              {\Var{G'}}
              {\Var{b}}}
          {\decom
              {\Var{G}}
              {\Var{t}}
              {\Var{C}}
              {\Var{t'}}
              {(\cnamep{\Var{x}}{\Var{p}})}
              {\Var{G'}}
              {\bunion{\Var{b}}{\{ (\Var{x}, \Var{C}) \}}}}\\
\end{mathpar}
\caption{Modified specification of decomposition.}
\flabel{decomp_spec}
\end{figure}

The first rule specifies the process of decomposition of a given term \Var{t},
when the pattern that describes the possible context is just \cholep{}. In 
that case, since such context only matches against \cholet{}, the term \Var{t}
is decomposed between a context that is just a single hole, and \Var{t} itself
as the sub-term. No binding is generated.

The second and third rules explain the decomposition of a list of terms 
\cconst{\Var{t}$_{hd}$}{\Var{t}$_{tl}$}, between a context that matches against a 
list of patterns \cconsp{\Var{p}$_{hd}$}{\Var{p}$_{tl}$}, and some sub-term.
In the second rule, the hole of the resulting context is pointing to somewhere
in the head of the list of terms. This information is indicated by the constructor
of the resulting context: \chdcont{\Var{C}}{\Var{t}$_{tl}$}, where \Var{C} is some
context that must match against pattern \Var{p}$_{hd}$, as indicated in the
premise of the inference rule. Indeed, recall that the context from the decomposition 
must match against pattern \cconsp{\Var{p}$_{hd}$}{\Var{p}$_{tl}$}. If this patterns 
is describing some context whose hole points to somewhere in the head of the list of terms,
it must be the case that the sub-pattern \Var{p}$_{hd}$ matches against some
context. Note that the whole premise is stating that the decomposition occurs in the
head of the list of terms (\Var{t}$_{hd}$), and the resulting sub-term is 
\Var{t'}$_{hd}$. Then, the side-condition from the inference rule states that the 
tail of the original input term, \Var{t}$_{tl}$, must match against the tail of the
list of patterns \Var{p}$_{tl}$. Finally, note that in the decomposition through 
sub-pattern \Var{p}$_{hd}$, and the matching against sub-pattern \Var{p}$_{tl}$,
the non-terminals of these patterns are interpreted in terms of productions from
the original grammar, \Var{G}. This is done since, in each case, we are operating
over a proper sub-term of the original input.

The third rule can be explained on the same basis as in the previous case, with the
sole difference that, now, the context from the resulting decomposition is pointing
to somewhere in the tail of the original list of terms. Note that, in both rules,
the resulting bindings are the disjoint union of bindings from the decomposition and
the matching step.

The remaining rules can also be understood in similar terms as with the previous rules,
the exception being the case of the \cinholept{} pattern. Note that this situation
corresponds to an original pattern of the form:
\begin{center}
\cinholep{(\cinholep{\Var{p}$_c$}{\Var{p}$_h$})}{\Var{p}$_{h'}$}
\end{center}
that we matched against some term \Var{t}. The semantics of this involves a first
step of decomposition of \Var{t} between some context that matches against
sub-pattern \cinholep{\Var{p}$_c$}{\Var{p}$_h$}, and some sub-term that matches
against sub-pattern \Var{p}$_{h'}$. In the rules from \fref{decomp_spec} for the 
case of pattern \cinholept{}, we are describing what it means, in this situations, 
that first step of decomposing \Var{t} in terms of a context that matches against 
pattern \cinholep{\Var{p}$_c$}{\Var{p}$_h$}. Since the whole pattern must
match against some context, it means that, both, \Var{p}$_c$ and \Var{p}$_h$, are
patterns describing contexts. Now, through the pattern \cinholep{\Var{p}$_c$}{\Var{p}$_h$}
we are decomposing again the context, a first part that should match
against \Var{p}$_c$, and a nested context (to be put within the hole of the
previous context) that matches against \Var{p}$_h$. This idea is expressed through
the premises of both inference rules for the case of the \cinholept{} pattern.
Note that, again, we distinguish the case where \Var{p}$_c$ produces an empty
context, from the case where it does not. The intention being to be able to
recognize whether we should interpret non-terminals from patterns through
the original grammar \Var{G} or the arbitrary grammar \Var{G'}.

The last piece of complexity of the rules for the \cinholept{} pattern resides
in the actual context that results from the decomposition. Here, the authors of
\redexk, expressed this context as the result of plugging one of the obtained
contexts within the other, denoted with the expression $\ccon{\Var{C}_c}{\Var{C}_h}$:
this represents the context obtained by plugging context $\Var{C}_h$ within the
hole of context $\Var{C}_c$, following the information contained in the constructor
of the context $\Var{C}_h$ to find its actual hole. For reasons of space we elude
this definition, though it presents no surprises.

\subsubsection{Matching and decomposition algorithm}
\label{sec:match_decom_alg}
We close this section presenting a simplified description of the matching and 
decomposition algorithm adapted for its mechanization in Coq. We remind the reader 
that this algorithm is just a modification of the one proposed for \redexk{}, 
in \cite{semcontext}.

Naturally, the actual mechanization is far more complex than what we present here. 
The intention is to provide the reader with a high-level view of the main ideas 
behind the mechanization. 

The previous specification of the algorithm cannot be used directly to derive
an actual effective procedure to compute matching and decomposition. In particular,
the rules for decomposition of lists of terms (second and third rules from 
\fref{decomp_spec}) do not suggest effective meanings to determine whether to
decompose on the head, and match on the tail, or vice versa. To solve this issue
(and the complexity problem that could arise from trying to naively perform both kind
of decomposition simultaneously), the algorithm developed for \redexk{}
performs matching and decomposition simultaneously but sharing intermediate results.
\begin{figure}
\begin{lstlisting}
Definition binding := prod var term.
Definition bindings := list binding.

Inductive decom_ev : term -> Set :=
  | empty_d_ev : forall (t : term), decom_ev t 
  | nonempty_d_ev : forall t (c : contxt) subt,
      {subt = t /\ c = hole_contxt_c} + {subterm_rel subt t} -> decom_ev t.

Inductive mtch_ev : term -> Set :=
  | mtch_pair : forall t, decom_ev t -> bindings -> mtch_ev t.

Definition mtch_powset_ev (t : term) := list (mtch_ev t).
\end{lstlisting}
\caption{Mechanization of decomposition and matching results.}
\label{fig:decom_match_impl}
\end{figure}

\paragraph{Supporting data-structures.}
In \fref{decom_match_impl} we show some of the implemented data-structures used
to represent the results returned by \redexk{}'s algorithm. The result of a matching/decomposition 
of a term \Var{t} (against some given pattern) will be represented through a value of type 
\lstl{mtch\_ev} \Var{t}. Naturally, making the type dependent on \Var{t} is done for future
soundness checking. The algorithm could return several values of this type, each one representing 
a possible match or a decomposition. We represent this several values through the list type
\lstl{mtch\_powset\_ev} \Var{t}.\footnote{Naturally, this allows for repeated values to occur in the 
result. This does not affect desired soundness properties.}

For reasons of brevity, when presenting the algorithm we will avoid the actual concrete syntax 
from our mechanization. A value of type \lstl{mtch\_ev} \Var{t} will be denoted as
$\mtchpair{}{\Var{d}}{\Var{b}}$, where \Var{d} is a value of type \lstl{decom_ev} \Var{t}
(explained below), and \Var{b} is a list of bindings (also shown in \fref{decom_match_impl}).
For reasons of brevity, we drop the dependence of the previous value on term \Var{t}, in the notation
used. We will maintain the same notation used so far for bindings. In particular, recall that an empty 
list of bindings is denoted as $\oslash$. For a value of the list type 
\lstl{mtch\_powset\_ev} \Var{t}, we will denote it decorating it with its dependence on the value
\Var{t}: $[\mtchpair{}{\Var{d}}{\Var{b}}, ...]_{\Var{t}}$

Values type \lstl{decom_ev} \Var{t} represent a decomposition of a given term \Var{t}, 
between a context and a sub-term. We make the type dependent on \Var{t} for soundness checking
purposes, and we include in the value some evidence of soundness of the decomposition: 
in particular, evidence showing that a sub-term \Var{subt} extracted in the 
decomposition is, either, \Var{t} itself (proof of type $\Var{subt} = \Var{t}$) or a proper 
sub-term of \Var{t} (proof of type \lstl{subterm\_rel} \Var{subt} \Var{t}). Recall that
\lstl{subterm\_rel} is our mechanization of relation $\subt{}{}$ (see \sref{input_consumption_sec}). 
Soundness properties about a context \Var{c} extracted in the decomposition are, either, 
embedded in the \lstl{decom_ev} value itself (\Var{c} \lstl{= hole_contxt_c}, when \lstl{subt = t}),
or they emanate from properties stated through the formal system that captures decomposition (note that 
this system does not explicitly specify properties about the sub-term extracted in the decomposition, 
but it does capture the context).

Since a value of type \lstl{mtch\_ev} \Var{t} could represent a single match or a single 
decomposition, we distinguish an actual match using an \empty{empty} decomposition \lstl{empty_d_ev} \Var{t}
(denoted as $\empdec{\Var{t}}$ or, simply, $\empdec{}$, when it is clear from context the
actual term \Var{t} upon which the value depends). A value of type \lstl{mtch\_ev} \Var{t} that actually represents 
a decomposition, will contain a value \lstl{decom_ev} \Var{t} of the form 
\lstl{nonempty\_d\_ev} \Var{t} \Var{C} \Var{subt} \Var{ev}, for a proper context \Var{C}, 
sub-term \Var{subt} and soundness evidence \Var{ev}. We will denote values constructed this
way as \nempdecev{\Var{t}}{\Var{C}}{\Var{subt}}{\Var{ev}}. For a proof of type 
$\mathtt{\{\Var{subt} = \Var{t} \wedge \Var{C} = hole\_contxt\_c\} + \{subterm\_rel \; \Var{subt} \;\Var{t}\}}$, when we can 
determine the actual disjunct proved we will indicate it with its type. For example,
if we know that what it actually holds is $\mathtt{subterm\_rel \; \Var{subt} \;\Var{t}}$,
we will write \nempdecev{\Var{t}}{\Var{C}}{\Var{subt}}{\mathtt{subterm\_rel \; \Var{subt} \;\Var{t}}}.
Also, in this context we will write just $\mathtt{\Var{subt} = \Var{t}}$, when what it holds is predicate 
$\mathtt{\Var{subt} = \Var{t} \wedge \Var{C} = hole\_contxt\_c}$.
Finally, when it is required to simplify the notation, and when it is clear from the context, we will allow us not 
to include information about dependence on the particular term \Var{t} that is being considered.

\paragraph{Matching and decomposition algorithm as a least-fixed-point.}
As is common practice in functional programming, we will capture the intended matching/decomposition 
algorithm as the least fixed-point of a \emph{generator function} or \emph{functional}. That is, we
will provide equations that actually capture a function that receives an \emph{approximation} of our intended 
algorithm, and uses it to return a better approximation. Provided that we can show that this generator 
function respects our well-founded relation (described in \sref{well_founded}), through Coq's \lstl{Fix} 
combinator we can get, in return, a function that is \emph{total} over the domain of that relation. Now, looking
at \lstl{Fix}'s implementation, we see that it defines a process that unfolds our generator function 
\emph{only} as much as needed to reach to a result, doing primitive recursion over the proof of 
accessibility (with respect to the provided well-founded relation) of the parameter upon which 
we are evaluating our generator.\footnote{For example, in \url{https://coq.inria.fr/refman/language/coq-library.html\#index-20}} 
So, this process is guaranteed to terminate and it can be shown that it emulates the behavior of 
the least fixed-point of our generator function.\footnote{If we can also prove that our generator function does not \emph{distinguish} extensionally equal approximations, we can also get a proof showing that through the \lstl{Fix} combinator we get a fixed-point of our generator function (lemma \lstl{Fix_eq} from Coq's standard library).
The fact that it is also the least fixed-point, could be proved on the basis of the way in which \lstl{Fix} operates.}

This fixed-point will be a function that captures so good our intended algorithm, that it cannot be 
\emph{improved} by our generator function: \ie it is a fixed-point and, even more, \emph{is} the 
intended algorithm.

Consider the following function type:
\begin{quote}

\begin{lstlisting}
 forall (g1 : grammar) (tpg1 : matching_tuple),
  (forall tpg2 : matching_tuple, matching_tuple_order g1 tpg2 tpg1 
                                  -> 
                                  mtch_powset_ev (matching_tuple_term tpg2)) 
   ->
   mtch_powset_ev (matching_tuple_term tpg1)
\end{lstlisting}
\end{quote}

It corresponds to a family of generator functions, $\mevbody$, parameterized over
grammars and tuples of terms and patterns, and that will \emph{improve} candidates
of matching/decomposition functions. 

In the previous type, 
\lstl{matching_tuple} is defined as the type \lstl{term} $\times$ \lstl{pat} $\times$ \lstl{grammar} and
\lstl{matching_tuple_order} is the mechanization of our well-founded relation
from \sref{well_founded}. Finally, for a given value \Var{tpg} \lstl{: matching_tuple},
\lstl{matching_tuple_term} $\Var{tpg}$ \lstl{: term} is just the projection of the first 
component of \Var{tpg}. Note that, for given \Var{G1} \lstl{: grammar}, tuple 
\Var{Tpg1} \lstl{: matching_tuple}, then, $\mevbody \; \Var{G1} \;\Var{Tpg1} $ has
type:
\begin{quote}
\lstinline{(forall tpg2 : matching_tuple, matching_tuple_order $\Var{G1}$ tpg2 $\Var{Tpg1}$}\\
\lstinline{                                ->}\\
\lstinline{                                mtch_powset_ev (matching_tuple_term tpg2))}\\
\lstinline{->}\\
\lstinline{mtch_powset_ev (matching_tuple_term $\Var{Tpg1}$)}
\end{quote}

Hence, $\mevbody \; \Var{G1} \;\Var{Tpg1} $ will be our intended generator function for grammar
\Var{G1} and matching tuple \Var{Tpg1}. It receives as a parameter a function that it will improve: 
the type of this parameter codifies the idea that it is a function that knows how to compute the matching/decomposition for any
tuple that is \emph{smaller} than \Var{Tpg1}, according to \lstl{matching_tuple_order} \Var{G1}.
Finally, in $\mevbody \; \Var{G1} \;\Var{Tpg1} $, \Var{G1} will represent the original 
grammar over which we want to compute the matching and decomposition indicated by the tuple
\Var{Tpg1}.

\begin{figure}

{\newcommand\hscons{1.7} 
  \newcommand\hsinh{1.7} 
  \begin{align*}
    &\mevbody (\_, \mtup{\holet}{\holet}{\_}, \_)
   & &= [\mtchpair
             {}
             {\nempdecev
                  {}
                  {\holet}
                  {\holet}
                  {\holet = \holet}}
             {\oslash}, 
         \mtchpair{}
                  {\empdec{}}
                          {\oslash}]_{\holet}\\
    &\mevbody (\_, \mtup{\Var{t}}{\holet}{\_}, \_)
   & &= [\mtchpair
             {}
             {\nempdecev
                  {}
                  {\holet}
                  {\Var{t}}
                  {\Var{t} = \Var{t}}}
             {\oslash}]_\Var{t}\\
    &\mevbody (\_, \mtup{\lit{a}}{\lit{a}}{\_}, \_)
   & &= [\mtchpair
             {}
             {\empdec {}}
             {\oslash}]_{\lit{a}}\\
    &\mevbody (\_, \mtup{\nil}{\nil}{\_}, \_)
   & &= [\mtchpair
             {}
             {\empdec {}}
             {\oslash}]_{\nil}\\
    &\mevbody (\Var{g}_1, 
      \mtup{\Var{t}}
           {\Var{p}}
           {\Var{g}_2}, \Var{M}_{ap}) 
   & &= [\mtchpair
             {}
             {\Var{d}}
             {\Var{b}} \; | \; \Var{d} \in \selectev{\Var{t}_{hd}}{\Var{d}_{hd}}
                                                   {\Var{t}_{tl}}{\Var{d}_{tl}}{\Var{t}}{\Var{sub}},\\
    &&&    \hspace{\hscons cm} \Var{sub} : \subterms{\Var{t}}{\Var{t}_{hd}}{\Var{t}_{tl}}\\
    &&&    \hspace{\hscons cm} \Var{b} = \bunion{\Var{b}_{hd}}{\Var{b}_{tl}},\\
    &&&    \hspace{\hscons cm} \mtchpair 
                                 {\Var{t}_{hd}}
                                 {\Var{d}_{hd}} 
                                 {\Var{b}_{hd}} \in \Var{M}_{ap} (\Var{tp}_{hd}, \Var{lt}_{hd}),\\
    &&&    \hspace{\hscons cm}  \mtchpair
                                 {\Var{t}_{tl}}
                                 {\Var{d}_{tl}}
                                 {\Var{b}_{tl}} \in \Var{M}_{ap} (\Var{tp}_{tl}, \Var{lt}_{tl}),\\
    &&&    \hspace{\hscons cm} \Var{lt}_{hd} : \lexprod{\Var{g}_1}{\Var{tp}_{hd}}{\Var{tp}_{cons}},\\
    &&&    \hspace{\hscons cm} \Var{lt}_{tl} :\lexprod{\Var{g}_1}{\Var{tp}_{tl}}{\Var{tp}_{cons}},\\
    &&&    \hspace{\hscons cm} \Var{tp}_{cons} = \mtup{\Var{t}}
                                                {\Var{p}}
                                                {\Var{g}_2},\\
    &&&    \hspace{\hscons cm} \Var{tp}_{hd} = \mtup{\Var{t}_{hd}}{\Var{p}_{hd}}{\Var{g}_1},\\
    &&&    \hspace{\hscons cm} \Var{tp}_{tl} = \mtup{\Var{t}_{tl}}{\Var{p}_{tl}}{\Var{g}_1}]_{\Var{t}},\\
    &&& \hspace{\hscons cm} with \; \Var{t} = \consp{\Var{t}_{hd}}{\Var{t}_{tl}}\\
    &&& \hspace{\hscons cm}\hspace{0.8cm} \Var{p} = \consp{\Var{p}_{hd}}{\Var{p}_{tl}}\\
    &\mevbody (\Var{g}_1, \mtup{\Var{t}}{\Var{p}}{\Var{g}_2}, \Var{M}_{ap}) 
   & &= [\mtchpair
             {}
             {\Var{d}}
             {\Var{b}} \; | \; \Var{d} = \combine{\Var{t}}{\Var{C}}{\Var{t}_c}{\Var{ev}}{\Var{d}_h},\\
    &&&     \hspace{\hsinh cm} \Var{b} = \bunion{\Var{b}_c}{\Var{b}_h},\\
    &&&    \hspace{\hsinh cm}  \mtchpair
                                 {\Var{t}_c}
                                 {\Var{d}_h}
                                 {\Var{b}_h} \in \Var{M}_{ap} (\Var{tp}_h, \; \Var{lt}_h),\\
    &&&    \hspace{\hsinh cm} \Var{lt}_h : \lexprod{\Var{g}_1}{\Var{tp}_h}{\Var{tp}_{inhole}},\\
    &&&    \hspace{\hsinh cm} \Var{tp}_h = \mtup{\Var{t}_c} 
                                                  {\Var{p}_h}
                                                  {\Var{g}_h},\\
    &&&    \hspace{\hsinh cm} \Var{g}_h \; \textrm{according to \fref{match_spec}},\\
    &&&    \hspace{\hsinh cm}  \mtchpair
                                 {\Var{t}}
                                 {\nempdecev{\Var{t}}{\Var{C}}{\Var{t}_c}{\Var{ev}}}
                                 {\Var{b}_{c}} \in \Var{M}_{ap} (\Var{tp}_c, \Var{lt}_c),\\
    &&&    \hspace{\hsinh cm} \Var{lt}_c : \lexprod{\Var{g}_1}{\Var{tp}_c}{\Var{tp}_{inhole}},\\
    &&&    \hspace{\hsinh cm} \Var{tp}_{inhole} = \mtup{\Var{t}} 
                                                  {\Var{p}}
                                                  {\Var{g}_2},\\
    &&&    \hspace{\hsinh cm} \Var{tp}_c = \mtup{\Var{t}} 
                                                  {\Var{p}_c}
                                                  {\Var{g}_2}]_{\Var{t}},\\
    &&& \hspace{\hscons cm} with \; \Var{p} = \inholep{\Var{p}_c}{\Var{p}_h}\\
    &\mevbody (\Var{g}_1, \mtup{\Var{t}}{\Var{p}}{\Var{g}_2}, \Var{M}_{ap}) 
   & &= [\mtchpair
             {}
             {\Var{d}}
             {\Var{b'}} \; | \;\Var{b'} = \bunion{\{ (\Var{x}, \named{\Var{t}}{\Var{d}}) \}}{\Var{b}},\\
    &&&    \hspace{\hsinh cm} \hspace{0.1cm} \mtchpair
                                             {}
                                             {\Var{d}}
                                             {\Var{b}} \in \Var{M}_{ap} (\Var{tp}_{p'}, \Var{lt}_{p'})),\\
    &&&    \hspace{\hsinh cm} \hspace{0.1cm} \Var{lt}_{p'} : \lexprod{\Var{g}_1}{\Var{tp}_{p'}}{\Var{tp}_{name}},\\
    &&&    \hspace{\hsinh cm} \hspace{0.1cm} \Var{tp}_{name} = \mtup{\Var{t}} 
                                                          {\Var{p}}
                                                          {\Var{g}_2},\\
    &&&    \hspace{\hsinh cm} \hspace{0.1cm} \Var{tp}_{p'} = \mtup{\Var{t}} 
                                                           {\Var{p'}}
                                                          {\Var{g}_2}]_{\Var{t}},\\
    &&& \hspace{\hscons cm} \hspace{0.1cm} with \; \Var{p} = \namep{\Var{x}}{\Var{p'}}\\
   &\mevbody (\Var{g}_1, \mtup{\Var{t}}{\ntp{n}}{\Var{g}_2}, \Var{M}_{ap}) 
   & &= [\mtchpair
             {}
             {\Var{d}}
             {\oslash} \; | \; \mtchpair
                                 {}
                                 {\Var{d}}
                                 {\Var{b}} \in \Var{M}_{ap} (\Var{tp}_{\Var{p}}, \Var{lt}_{\Var{p}}),\\
    &&&    \hspace{\hsinh cm} \hspace{0.2cm} \Var{lt}_{\Var{p}} : \lexprod{\Var{g}_1}{\Var{tp}_{\Var{p}}}{\Var{tp}_{\Var{n}}},\\
    &&&    \hspace{\hsinh cm} \hspace{0.2cm} \Var{tp}_{\Var{n}} = \mtup{\Var{t}} 
                                                  {\ntp{\Var{n}}}
                                                  {\Var{g}_2},\\
    &&&    \hspace{\hsinh cm} \hspace{0.2cm} \Var{tp}_{\Var{p}} = \mtup{\Var{t}} 
                                                  {\Var{p}}
                                                  {\delprod{\Var{g}_2}{\Var{n}}{\Var{p}}},\\
    &&&    \hspace{\hsinh cm} \hspace{0.2cm} \Var{p} \in \Var{G}(\Var{n})]_{\Var{t}}
  \end{align*}
}
\caption{Generator function for the matching and decomposition algorithm.}
\flabel{match_alg}
\end{figure}

\fref{match_alg} shows the equations that capture \mevbody. For reasons of space, we describe
terms and patterns avoiding the more verbose concrete syntax of our mechanization. Also,
we will employ the same syntax for terms and patterns, resorting to context for disambiguation.
In general the syntax is self-explanatory: for example \holet{} represents a hole term or 
pattern (depending on context), \lit{a} represents a literal term or pattern, etc.

The first 4 equations of \fref{match_alg} can be understood by comparison with the 
specifications of matching (\fref{match_spec}) and decomposition (\fref{decomp_spec}). 
For example, the first equation explains the matching and decomposition of a term 
\holet{} against a pattern \holet{}. Note that the second inference rule of matching
specifies that such term matches against such pattern, producing no bindings: this is
represented by the pair \mtchpair{}{\empdec{}}{\oslash}, in the result captured in the 
first equation. Also, the first rule of decomposition specifies that given some
term \Var{t}, it can be decomposed between a single hole (that matches against 
pattern \holet{}) and \Var{t} itself, producing no bindings.
Assuming $\Var{t} = \holet$, the pair \mtchpair{}{\nempdecev{}{\holet}{\holet}{\holet = \holet}}{\oslash}
shown in the result captured in the first equation represents the described situation:
the context extracted is a single hole, and the sub-term is $\Var{t} = \holet$
itself. Note that we also indicate the type of the actual proof contained in the piece
of evidence of soundness of the decomposition: in this case, a proof showing
that the sub-term extracted is actually equal to the original term itself 
($\holet = \holet$). In the described pairs we dropped the mention to the actual term
whose types depend on, but we do mention it for the whole list type containing the 
previous pairs. Finally, we do not name the parameters to \mevbody{} that are not
mentioned in the right-hand side of the equations, using a wildcard ``\_'' instead.

The fifth equation explains the matching and/or decomposition of a list of terms
($\consp{\Var{t}_{hd}}{\Var{t}_{tl}}$) against a list of patterns ($\consp{\Var{p}_{hd}}{\Var{p}_{tl}}$). 
We describe by comprehension the list of results. Note 
that, to explain this case, we need to consider the approximation function 
$\Var{M}_{ap}$ that \mevbody{} receives as its last parameter. We begin by using
$\Var{M}_{ap}$ to compute matching and decomposition for \emph{smaller} tuples:
$\Var{tp}_{hd} = \mtup{\Var{t}_{hd}}{\Var{p}_{hd}}{\Var{g}_1}$
and $\Var{tp}_{tl} = \mtup{\Var{t}_{tl}}{\Var{p}_{tl}}{\Var{g}_1}$. Note that,
given that these tuples represent a matching/decomposition over a proper sub-term 
of the input term, we consider the original grammar $\Var{g}_1$ (first parameter of 
\mevbody). In order to be able to fully evaluate $\Var{M}_{ap}$, we need to
build proofs \Var{lt}$_{hd}$ and \Var{lt}$_{tl}$ of type 
$\lexprod{\Var{g}_1}{\Var{tp}_{hd}}{\Var{tp}_{cons}}$ and 
$\lexprod{\Var{g}_1}{\Var{tp}_{tl}}{\Var{tp}_{cons}}$, respectively, 
where $\Var{tp}_{cons}$ is the original tuple over which we evaluate \mevbody. 
Then, for each value
of type \lstl{mtch_ev} \Var{t}$_{hd}$ and \lstl{mtch_ev} \Var{t}$_{tl}$ of the
results obtained from evaluating $\Var{M}_{ap}$, the algorithm inspect if they
are decompositions or not, and if it is possible to combine these results,
using the helper function \textsf{select}. 

The original \textsf{select} helper function
from \redexk{} receives as parameters \Var{t}$_{hd}$, \Var{d}$_{hd}$, \Var{t}$_{tl}$
and \Var{d}$_{tl}$. It analyses \Var{d}$_{hd}$ and \Var{d}$_{tl}$: if none of them
represent actual decompositions (\ie they are values of the form \empdec{}),
then the whole operation will be considered just a matching of the original 
list of terms (rule for matching of \cconstt, \fref{match_spec}) and \textsf{select} 
must build an \emph{empty} decomposition of the proper type to represent this:
\lstl{decom_ev} (\Var{$\consp{\Var{t}_{hd}}{\Var{t}_{tl}}$}). If only \Var{d}$_{hd}$
is a decomposition, of the form \nempdecev{\Var{t}_{hd}}{\Var{C}}{\Var{t}_{hd'}}{\Var{ev}_{hd}},
then the whole operation is interpreted as a decomposition of the original list of 
terms on the head of the list (first rule of decomposition for \cconstt, 
\fref{decomp_spec}). In that case, \textsf{select} builds a value of type 
\lstl{decom_ev} ($\consp{\Var{t}_{hd}}{\Var{t}_{tl}}$), of the form 
\nempdecev{\consp{\Var{t}_{hd}}{\Var{t}_{tl}}}{\chdcont{\Var{C}}{\Var{t}$_{tl}$}}{\Var{t}_{hd'}}{\Var{ev}_{\consp{\Var{t}_{hd}}{\Var{t}_{tl}}}}. 
Observe the correspondence between, on the one hand, the context and the sub-term from
this decomposition, and, on the other hand, the context and sub-term specified in the 
first rule of decomposition for \cconstt, \fref{decomp_spec}. Finally, for the
mechanization of \textsf{select} to be able to build the required soundness proofs
of decomposition (for \lstl{decom_ev}), we need to provide to it the original
list of terms, and evidence \Var{sub} showing that \Var{t}$_{hd}$ and \Var{t}$_{tl}$ are the
actual head and tail of the original input term 
($\Var{sub} : \subterms{\Var{t}}{\Var{t}_{hd}}{\Var{t}_{tl}}$).

The remaining equations can be understood on the same basis as the previous one,
requiring only some explanation the equations for the patterns \inholet{} and \namet{}:
in the first case, the auxiliary function \textsf{combine} helps in deciding if the
result is a decomposition against pattern \inholet{}, or if it is just a match against
said pattern, depending on whether \Var{d}$_h$ is a decomposition or not; in the case
of the \namet{} pattern, the auxiliary function \textsf{named} plays a similar role as
the previous one: it helps in deciding if the result is a decomposition or a matching,
in order to make the binding refer to the context extracted (case of \namet{} pattern
in \fref{decomp_spec}) or the actual input term (case of \namet{} pattern
in \fref{match_spec}), respectively.

Finally, as mentioned at the beginning of this section, we define the desired 
matching/decomposition algorithm, $\mev$, as the least fixed-point of the previous 
generator function. We show in \fref{mev} its Coq definition.

\begin{figure}
\begin{lstlisting}
Definition Mev (g : grammar) (tup : matching_tuple) :  
mtch_powset_ev (matching_tuple_term tup) :=
(Fix
 (matching_tuple_order_well_founded g)
 (* dependent range type of the function that we are building *)
 (fun tup : matching_tuple => mtch_powset_ev (matching_tuple_term tup))
 (* generator function *)
 (Mev_gen g))
tup.
\end{lstlisting}
\caption{Definition of $\mev$ in Coq.}
\label{fig:mev}
\end{figure}

\subsubsection{Semantics for context-sensitive reduction rules}

The last component of \redexk{} consists in a semantics for context-sensitive 
reduction rules, with which we define semantics relations in Redex (for example, 
the semantics rules presented in \sref{red_semantics} to capture $\beta$-contractions).
The proposed semantics makes use of the introduced notion of matching, to define
a new formal system that explains what it means for a given term to be \emph{reduced},
following a given semantics rule. 

We have mechanized the previous formal system, though, for reasons of space, we do not
introduce it here in detail. Its mechanization does not require the development of new
concepts or the use of complex tools, and it relies, heavily, on the notion of
matching previously presented. The reader is invited to look at the mechanization of
this formal system, in module \textsf{reduction.v} of the source code accompanying this
paper.


\section{Soundness of matching}
\label{sec:soundness}
In the original paper of \redexk{} it is proved the expected correspondence between 
the presented algorithm and its specification. In our mechanization we reproduced those 
results, for the least-fixed-point of \mevbody \; \Var{g} (\Var{t}, \Var{p}, \Var{g'}),
for arbitrary \Var{g}, \Var{t}, \Var{p} and \Var{g'}. Naturally, for a given
grammar \Var{g}, the original intention of matching and decomposition corresponds to 
the least-fixed-point of \mevbody \; \Var{g} (\Var{t}, \Var{p}, \Var{g}). In what
follows, \Var{M\_ev} \Var{g} (\Var{t}, \Var{p}, \Var{g'}) will represent the 
least-fixed-point of \mevbody \; \Var{g} (\Var{t}, \Var{p}, \Var{g'}).

With respect to the soundness checks of the mechanized version of the matching/decomposition 
algorithm, we were able to implement a proof of the completeness of the process,
with respect to its specification. We show its statement in \fref{completeness_mev}.

\begin{figure}
\begin{theorem}[Completeness of \Var{M\_ev}]
\begin{lstlisting}
Theorem completeness_M_ev : 

forall G1 G2 p t sub_t b C,
 (G1 |- t : p, G2 | b              -> In (mtch_pair t (empty_d_ev t) b)
                                         (M_ev G1 (t, (p, G2))))
 /\
 (G1 |- t1 = C [ t2 ] : p , G2 | b -> exists (ev_decom :       {sub_t = t} 
                                                                   + 
                                                         {subterm_rel sub_t t}),
                                       In (mtch_pair t (nonempty_d_ev t C sub_t 
                                                        ev_decom) b)
                                       (M_ev G1 (t, (p, G2)))).

\end{lstlisting}
\end{theorem}
\caption{Mechanization of the proof of completeness of \Var{M\_ev}: statement.}
\flabel{completeness_mev}
\end{figure}

Note that we represent and manipulate results returned from \Var{M\_ev} through
Coq's standard library implementation of lists. Also, the shape of the tuples of terms, patterns
and grammars, is the result of the way in which we build our lexicographic product: the
product between a relation with domain \lstl{term}, and a relation with domain
\lstl{pat} $\times$ \lstl{grammar}.

\begin{figure}
\begin{theorem}[Soundness of \Var{M\_ev}]
\begin{lstlisting}
Theorem soundness_M_ev : 

forall G1 G2 p t sub_t b C,
 (In (mtch_pair t (empty_d_ev t) b)
     (M_ev G1 (t, (p, G2)))            ->    G1 |- t : p, G2 | b)
 /\
 (exists (ev_decom :     {sub_t = t} 
                              + 
                     {subterm_rel sub_t t}),
    In (mtch_pair t (nonempty_d_ev t C sub_t 
                     ev_decom) b)
    (M_ev G1 (t, (p, G2)))             ->    G1 |- t1 = C [ t2 ] : p , G2 | b).

\end{lstlisting}
\end{theorem}
\caption{Mechanization of the proof of soundness of \Var{M\_ev}: statement.}
\flabel{soundness_mev}
\end{figure}

The converse, the soundness of the process with respect to its specification, was
also mechanized, and we show its statement in \fref{soundness_mev}. The proofs
present no surprises. Since we have a well-founded recursion over the tuples
from \lstl{term} $\times$ \lstl{pat} $\times$ \lstl{grammar}, we also have an
induction principle to reason over them. This is useful to prove soundness.
Completeness can be proved by \emph{rule induction} on the evidences of
match and decomposition.

\begin{figure}
\begin{lstlisting}
Definition pat_grammar_evolution_trans := 
  clos_trans_1n (pat * grammar) pat_grammar_evolution.

Definition non_left_recursive_grammar :=
  forall (p : pat) (g1 g2 : grammar), 
   not (pat_grammar_evolution_trans (p, g1) (p, g2)).
\end{lstlisting}
\caption{Non-left-recursive grammars.}
\flabel{non_left}
\end{figure}

We also verified the correspondence between our mechanized specification of 
matching and decomposition, and the original formal systems from the paper.
In doing so, we needed to prove, first, some soundness results about our
own formal systems. On the one hand, we needed to verify that our replacement
of grammars during the phase of non-consumption of input is actually sound.
In doing so, we required an actual formalization of our assumptions about the
grammars under consideration: that is, that they are non-left-recursive.
We show its definition in \fref{non_left}. Note that \lstl{pat_grammar_evolution}
is just the mechanized version of the relation \ninpconsordsym, that models
the phase of non consumption of input, during matching/decomposition. The
definition shown in \fref{non_left} is just an adaptation of the proposal 
of the authors of \redexk{}.

\begin{figure}
\begin{lemma}[Soundness of replacement of grammars]
\begin{lstlisting}
Lemma non_left_g_rm_sound : 

forall g1 g2 t n p (proof_in : prod_in_g (n, p) g2) b,
  non_left_recursive_grammar ->

  (g1 |- t : p, remove_prod (n, p) g2 proof_in | b
                     <->
              g1 |- t : p, g2 | b)
   
   /\
     
  (forall (C : contxt) (subt : term),
    g1 |- t = C[subt] : p, remove_prod (n, p) g2 proof_in | b
                         <->
               g1 |- t = C[subt] : p, g2 | b).
\end{lstlisting}
\end{lemma}
\caption{Mechanization of the proof of soundness of our manipulation of grammars: statement.}
\flabel{soundness_grammars}
\end{figure}

We make use of \lstl{non_left_recursive_grammar} to strengthen our 
assumptions, when verifying the soundness of our replacement of grammars. 
Our formal statement of soundness consists in specifying that, if we have already used some 
production (for example, when matching against a pattern \ntp{\Var{n}} and after
looking for some production of \Var{n} in the grammar under consideration), 
removing it from the grammar does not impede us to complete a proof of matching or 
decomposition. We show its statement in \fref{soundness_grammars}. In the
statement, \lstl{prod_in_g (n, p) g2} is a type (of sort \lstl{Prop}) that represents
proofs showing that \lstl{(n,p)} is actually a production from \lstl{g2}. Finally,
\lstl{remove_prod (n, p) g2 proof_in} builds a new grammar that contains the same
productions as \lstl{g2}, except for \lstl{(n, p)}.

\begin{figure}
\begin{lemma}[Soundness of decomposition]
\begin{lstlisting}
Lemma subterm_rel_characterization :

forall G p G' t C t' b,
 G |- t = C [t'] : p, G' | b -> subterm_rel t' t \/ (t = t' /\ C = hole__t).
\end{lstlisting}
\end{lemma}
\caption{Mechanization of the proof of soundness of decomposition: statement}
\flabel{soundness_decom}
\end{figure}

In addition, we needed to verify that our formal system to specify decomposition,
actually help us to build proofs about meaningful statements: in particular, 
if I can build a proof for a judgment like \lstl{G |- t = C [t'] : p, G'}, 
then it must happen that \lstl{t'} is a proper sub-term of \lstl{t}, or, 
if \lstl{t = t'}, then, the context \lstl{C} is simply a hole. The statement
of our mechanized proof for this result is shown in \fref{soundness_decom}.

\begin{figure}
\begin{theorem}[Completeness of our formal systems]
\begin{lstlisting}
Theorem from_orig :

  forall G t p b,
   non_left_recursive_grammar ->
   G |- t : p | b ->
   G |- t : p, G | b

  with from_orig_decomp :

   forall G C t1 t2 p b,
   non_left_recursive_grammar ->
   G |- t1 = C [ t2 ] : p | b -> G |- t1 = C [ t2 ] : p , G | b.
\end{lstlisting}
\end{theorem}
\caption{Mechanization of the proof of completeness of our formal systems: statement.}
\flabel{complet_formal}
\end{figure}

With the obtained tools, we were able to tackle the formal verification of
correspondence between our formal system and the original ones from \redexk.
\fref{complet_formal} shows the statement of our completeness result: under
the assumption that grammars are non-left-recursive, and given a judgment that
can be proved in one of the original formal systems, we can reproduce the
same result within our formal systems. In particular, if we can prove
\lstl{G |- t : p | b}, that is, a proof of matching under the original formal
system, within our formal system we can prove \lstl{G |- t : p, G | b}. Note that,
as expected, we begin interpreting the non-terminals from the pattern using
the same original grammar \Var{G}. A similar result is obtained for the formal
systems specifying decomposition.

\begin{figure}
\begin{lstlisting}
  Axiom gleq : grammar -> grammar -> Prop.
  Axiom gleq_refl : forall G, gleq G G.
  Axiom gleq_trans : forall G G' G'', gleq G G' -> gleq G' G'' -> gleq G G''.
  Axiom gleq_weakening : forall {G G' p}, gleq G' G -> prod_in_g p G' -> prod_in_g p G.
  Axiom gleq_remove: forall p G pf, gleq (remove_prod p G pf) G.
\end{lstlisting}
\caption{Mechanization of the proof of soundness of our formal systems: axiomatization of \lstl{gleq}.}
\flabel{gleq}
\end{figure}

Finally, to prove soundness of our specification, with respect to 
the original formal systems, we need to reduce the spectrum of possible grammars
over which we begin interpreting the non-terminals of the pattern. In particular,
to reproduce, with the original formal system, a proof about a judgment
\lstl{G |- t : p, G' | b}, we cannot allow for \lstl{G'} to be \emph{any} grammar.
We will be limited only to grammars that contain some, or all, of the productions
from the original grammar \lstl{G}. We formalize this concept through a relation
\lstl{gleq}, whose properties we present in \fref{gleq}. Note that the axioms
only ask for \lstl{gleq} to be reflexive, transitive and to express the idea
that, if we can prove \lstl{gleq G' G}, then, every production from \lstl{G'}
is also in \lstl{G}.

\begin{figure}
\begin{theorem}[Soundness of our formal systems]
\begin{lstlisting}
Theorem to_orig : 

forall G G' t p b,
   gleq G' G ->
   G |- t : p, G' | b -> G |- t : p | b

  with to_orig_decomp :

   forall G G' C t1 t2 p b,
   gleq G' G ->
   G |- t1 = C [ t2 ] : p , G' | b ->
   G |- t1 = C [ t2 ] : p | b
\end{lstlisting}
\end{theorem}
\caption{Mechanization of the proof of soundness of our formal systems: statement.}
\flabel{soundness_formal}
\end{figure}

With the previous tool we can, finally, tackle the expected prove of the statement showing
that, if we can conclude something using our formal system, we can reach to a similar conclusion
using the original specification. We show its statement in \fref{soundness_formal}.

A natural consequences of this results is that, if \Var{M} is the original 
matching/decomposition function from \redexk, and \Var{M\_no\_ev} is a function defined
in terms of \Var{M\_ev}, that removes possible repeated results and every piece of
type-dependency and soundness evidence embedded in the results of \Var{M\_ev}, then,
we have the following correspondence:
\begin{center}
$\Var{M} \; (\Var{g}, \Var{p}, \Var{t}) = \Var{M\_no\_ev} \; (\Var{g}, \Var{p}, \Var{t})$,
\end{center}

which is the expected correspondence with the original formalization of \redexk{}.


\section{Related work}
\label{sec:related_work}

\paragraph{Coq libraries for reduction-semantics and related concepts.}
\col{}~\citep{blanqui_koprowski_2011} is a mechanization in Coq of the theory of
well-founded rewriting relations over the set of first-order terms, 
applied to the automatic verification of termination certificates. It presents
a formalization, in a dependently-typed fashion, of several fundamental concepts
of rewriting theory, and the mechanization of several results and techniques
used by termination provers. Its notion of terms includes first-order terms with 
symbols of fixed and varyadic arity, strings, and simply typed lambda terms. 
\col{} does not implement a notion of a language of patterns offering support for 
context-sensitive restrictions, something that is ubiquitous in a Redex mechanization,
to define semantics rules, formal systems and meta-functions over the terms of
a given language. Also, Redex is not focused on well-founded rewritting relations,
but, rather, in arbitrary relations over terms of a language. In order to use 
\col{} to \emph{explain} Redex, it would require a considerable amount of work,
extending and/or modifying \col{}, to be able to encode the semantics of Redex's 
language of patterns. In doing so, the user that developed a model in Redex and
tries to compile it into Coq using our tool, would be forced to work over
a mechanization of the model, in Coq, that does not provide a direct explanation of 
its semantics (in terms of Redex's own semantics) but rather, through an encoding over 
\col{}.


Sieczkowski et. al present in \citet{auto_der} an implementation in Coq of the technique of
\emph{refocusing}, with which it is possible to extract abstract machines from
a specification of a reduction semantics. The derivation method is proved correct, 
in Coq, and the final product is a generic framework that can be used to
obtain interpreters (in terms of abstract machines), from a given reduction 
semantics that satisfies certain characteristics. In order to characterize a reduction 
semantics that can be \emph{automatically refocused} (\ie transformed into a
traditional abstract machine), the authors provide an axiomatization capturing
the sufficient conditions. Hence, the focus is put in allowing the representation
of certain class of reduction semantics (in particular, deterministic models for which 
refocusing is possible), rather than allowing for the mechanization of arbitrary models 
(even non-deterministic semantics), as is the case with Redex. Nonetheless, future development of our tool 
could take advantage of this library, since testing of Redex's models that are proved 
to be deterministic could make use of an optimization as refocusing, to extract interpreters 
that run efficiently, in comparison with the expensive computation model of reduction semantics.

\emph{Matching logic} is a formalism, useful to specify logical systems and their properties, 
that has at its hearth a notion of patterns and pattern matching. In \citet{matching_logic} it is
presented a mechanization in Coq of a version of this meta-language, including its syntax, its
notion of semantics, formal system and verified soundness results. In particular, its syntax defines a language of
\emph{patterns}, whose semantics is defined in terms of the set of elements (from a given model)
that match against this pattern. In that sense, since the domain of interpretation of the phrases 
of this language is the powerset of elements of the given model, matching logic is considered a 
multi-valued logic. Redex could be explained as a matching logic, with formulas that represent 
Redex's patterns to capture languages and relations, and whose model refer to the terms (or structures
containing terms) that match against these patterns. While this representation of Redex could be of 
interest for the purpose of studying the underlying semantics of Redex, this is not satisfactory for 
the purpose of providing the user with a direct explanation in Coq of her/his mechanization in Redex.

\section{Conclusion}
\label{sec:conclusion}

We adapted \redexk{}~\citep{semcontext} to be able to mechanize it into Coq. In particular,
we obtained a primitive recursive expression of its matching algorithm; we introduced 
modifications to its language of terms and patterns, to better adapt it to the future inclusion of features
of Redex absent in \redexk{}; we reproduced the soundness results shown in \citet{semcontext},
but adapted to our mechanization, while also verifying the expected correspondence between our
adapted formal systems, that capture matching and decomposition, and the originals from the
cited work.

This first iteration enables a plethora of future opportunities for improvement, both,
in pursuing a faithful and complete representation of Redex features into Coq, and in
improving the possibilities offered by Redex for the development of semantics models.

\paragraph{Extending the capabilities of our mechanized model.}
A natural next step in our development could consist in the addition of automatic routines
to transpile a Redex model into an equivalent model in Coq. Also, extending the language
with capabilities of Redex absent in \redexk{} would be of vital importance to allow our tool
to be of practical use. Our proposed modification for the language of patterns and terms, 
already implemented in this first iteration, enables us to easily include Redex's Kleene closure
of patterns. This could be a reasonable next step in increasing the set of Redex's features
captured by our mechanization.

Finally, another major update of the model would be the addition of typing annotations into the 
language of patterns, to automatically check well-formedness of the definitions 
of a given model. Naturally, this would come at the cost of limiting the possible Redex models 
that can be transpiled into Coq, with our tool. This is the usual tension between 
allowing expressivity of a language or enforcing well-formedness of the things that we can 
say with the language.

\paragraph{Further development of decidability results.}
While the main results shown in the present paper are related with the mechanization
of \redexk{} in Coq, we already mentioned our interest in pursuing the development of
a theory of decidability about a model in Redex. This, to offer to the user a
set of tactics and general tools to build decision procedures for the properties
that she/he needs to study about a given model. While in the present mechanization
we have already make explicit some decidability constraints, about the model being
mechanized, this restrictions are actually completely expected for a model in Redex
(\eg decidability of definitional equalities between atomic elements of the language).

Studying the \emph{intersection of languages} problem, adapted to the kind of grammars
that can be expressed in Redex, could be of importance to our development. The recognition 
of equivalence between regular expressions (RE) has been already studied, with several 
approaches proposed. The classical approach \citep{hopcroft} being representing each RE through their
corresponding finite automatons, and studying the resulting product automaton.
With regard to context-free grammars (CFG), the general problem of 
deciding equivalence between 2 CFG is known to be non-decidable \cite{hopcroft}.
However, there are simpler problems known to be decidable. In
\citet{cfg_intersection}, Nederhof et. al study the problem of deciding
intersection between a CFG and a non-recursive CFG. They show that the problem
is decidable and PSPACE-complete. They also show that the problem remains
decidable when generalized to several CFG, from which some of them are
non-recursive. Recognizing restricted grammars, in Redex style, with good decidability
properties could be reasonable next in pursuing these studies. 

\paragraph{Solving efficiency issues when testing within Redex.}
As we already mentioned in the previous section, there is already work done in Coq
implementing tools to extract an interpreter based on abstract machines, from a given
deterministic reduction semantics model~\citep{auto_der}. This capability could
solve the well-known performance problem in Redex, when trying to use a given mechanized
semantics relation to reduce a term, from within the tool. Even more, this
solution to the problem comes with the added benefit of a certified correspondence
between both interpreters (the original, in terms of reduction semantics,
and the extracted, in terms of a classical abstract machine), something that lacks 
other approaches, like re-writing by hand the whole model in another language.

\bibliographystyle{./jar/spbasic}
\bibliography{references}
\end{document}